\documentclass[twocolumn]{webofc}

\usepackage[varg]{txfonts}   
\usepackage[titletoc]{appendix}
\usepackage{color}
\newcommand{\be}{\begin{equation}}
\newcommand{\ee}{\end{equation}}
\newcommand{\bea}{\begin{eqnarray}}
\newcommand{\eea}{\end{eqnarray}}
\newcommand{\lf}{\left}
\newcommand{\rg}{\right}

\begin{document}
\title{Photo- and electro-production of narrow exotic states}
%
\subtitle{from light quarks to charm {and up} to bottom}

\author{\firstname{Xu} \lastname{Cao}\fnsep\thanks{\email{caoxu@impcas.ac.cn}}
}

\institute{Institute of Modern Physics, Chinese Academy of Sciences, Lanzhou 730000, China
\and
           University of Chinese Academy of Sciences, Beijing 100049, China
\and
           Research Center for Hadron and CSR Physics, Lanzhou University and Institute of Modern Physics of CAS, Lanzhou 730000, China
          }

\abstract{%
  Accessing a full image of the inner content of hadrons represents a central endeavour of modern particle physics, with the main scientific motivation to investigate the strong interaction binding the visible matter. On the one hand, the structure of known exotic candidates is a fundamental open issue addressed widely by scientists. On the other hand, looking for new states of exotic nature is a central component for theoretical and experimental efforts from electron-positron machine and electron accelerator with fixed target to heavy ion and electron-ion colliders. In this article we present a succinct short overview of the attempt to search for exotic narrow $N^*$ and $Z$ states containing light quarks only or also charm, and its connotation for bottom regions (the latter two are also known as $P_c$ ($Z_c$) and $P_b$ ($Z_b$) states, respectively in {the literature}). We address the effort of searching for exotic narrow $N^*$ and $Z$ states in light quark sector. We focus on recent progress in searching for signal of $P_c$ and $Z_c$ states photoproduction and its implication into the $P_b$ and $Z_b$ photoproduction and their decay properties. We also discuss future perspectives for the field in electron-ion colliders, a good place to disentangle the nature of some of these states and investigate some other enlightening topics including QCD trace anomaly and quarkonium-nucleon scattering length.
}
\maketitle
\section{Introduction}
\label{sec:intro}

Since the birth of quark model~\cite{GellMann:1964nj,Zweig:1964}, the multi-quark states which do not fit into standard picture of baryon containing three quarks or meson containing a pair of quark-antiquark, became of interest both for experimentalists and theorists.
The study of these unconventional states would deepen our understanding of non-perturbative Quantum Chromodynamics {QCD}, with the ultimate goal to reach a comprehension of properties of strong interaction which bind the visible matter.

The mesons with exotic quantum numbers, which are not allowed by two quark structure, are good candidates of muti-quark states \cite{Klempt:2007cp}.
The COMPASS collaboration have performed a full partial wave analysis of multi-particle final states with the aim of searching for the $\pi_1$ with quantum numbers $1^{-+}$ forbidden in quark model~\cite{Ketzer:2019wmd,COMPASS:2014vkj}.
The JPAC Collaboration claimed just recently solid evidence for the existence of $\pi_1$(1600) as lightest hybrid meson after tremendous efforts on pole analysis~\cite{Rodas:2018owy}, while another $\pi_1$ states with a lower mass of around 1400 MeV was not confirmed.
The width of $\pi_1$(1600) was determined to be very wide, namely around 500 MeV.
Its isoscalar partner candidate, $\eta_1$(1855) with a width of around 200 MeV, was just recently found by BESIII \cite{BESIII:2022riz,BESIII:2022iwi}.
The isovector partner of another broad exotic candidate $f_0$(1700) was observed as
$a_0$(1700) by BaBar \cite{BaBar:2021fkz} and $a_0$(1817) by BESIII \cite{BESIII:2022npc,BESIII:2021anf}.
Compared {to} the broad states, the narrow states in {the} light quark sector are much easier to be established as exotica if they are really in existence.
The COMPASS collaboration also ascribed a narrow resonancelike structure around 1400 MeV coupling to $f_0(980) \pi$ with an isovector axial-vector nature, the $a_1(1420)$, {caused by} triangle singularity (TS) as the origin of a genuine three-body effect \cite{COMPASS:2015kdx,COMPASS:2020yhb}, in line with previous theoretical calculations \cite{Mikhasenko:2015oxp,Aceti:2016yeb}.
Several models predicted charged $Z_s$ {states} (or $T_\phi$ in naming scheme of LHCb collaboration \cite{Gershon:2022xnn}) consisting of $u\bar{d}s\bar{s}$, close to $K \bar{K}^*$ or $K^* \bar{K}^*$ threshold~\cite{Chen:2011cj}. However, the BESIII Collaboration did not find their signals in the $\phi \pi$ spectrum of $e^+e^-\rightarrow\phi\pi\pi$~\cite{Ablikim:2018ofc}.

In the baryon sector, the famous $\Theta^+$ with the component $uddu \bar{s}$ disappeared in the large statistical $n K^+$ and $p K_S^0$ spectrum~\cite{Schumacher:2005wu}.
Recently, both constituent quark models~\cite{Huang:2018ehi,Liu:2018nse} and the models considering the QCD van der Waals force~\cite{Gao:2000az,He:2018plt} predicted the hidden-strangeness pentaquark $P_s$ (or $P^N_\phi$) states with $qqqs\bar{s}$ component ($q$: light $quark$).
The QCD sum rule \cite{Yang:2022uot} and unitary coupled-channel approximation \cite{Sun:2022cxf} did not exclude the existence of the $\phi p$ bound state.
Attractive $\phi p$ interaction was favored by correlation function from Lattice QCD calculation \cite{Lyu:2022imf} and ALICE measurement \cite{ALICE:2021cpv}.
The $\Lambda_c^+\to \phi p \pi^0$ was shown to be not a good choice for the search of $P_s$ due to the presence of triangle singularities and the tiny phase space {predicted by theoy}~\cite{Xie:2017mbe}.
As a matter of fact, Belle Collaboration showed no evident signal of resonance in its $\phi p$ spectrum ~\cite{Pal:2017ypp}.
Furthermore, no sharp peak of $P_s$ was found in near-threshold total cross section of $\gamma p\to \phi p$, but a non-monotonic structure, found in the differential cross section by LEPS Collaboration~\cite{Mibe:2005er}, would imply a very wide ($\sim$ 500 MeV) states $N^*$(2100)~\cite{Kiswandhi:2010ub,Kiswandhi:2011cq}.
This state, together with $N^*$(1875), were proposed to be exotic baryons~\cite{He:2017aps}. Alternative assignment was argued to be $N^*$(1875) and $N^*$(2080) (labeled as $N^*$(2120) in present version of PDG~\cite{ParticleDataGroup:2022pth}), which are close to $K \Sigma^*$ and $K^* \Sigma$ thresholds, respectively~\cite{Lin:2018kcc}.
However, these states decay fast to their ground state --- nucleon, so their widths are all bigger than 100~MeV.
As a result, it needs further effort to unambiguously {establish them as exotic states}.
On the other hand, the narrow states can be hardly incorporated into the traditional spectrum.
For instance, the $N^*$ and $\Delta^*$ spectrum \cite{Thiel:2022xtb}, {consisting} of states with excitation of internal {degrees of} freedom of nucleon and $\Delta$, must be {wider} than 100 MeV because of their strong coupling to $\pi N$, $\eta N$ and $\pi \pi N$, $etc$. So the existence of narrow nucleon resonances would serve as excellent evidence of exotica.

A renowned narrow structure at around 1680 MeV close to $K \Sigma$ threshold was found by Graal group in $\eta$ photoproduction off neutron~\cite{Kuznetsov:2006kt} and confirmed by many other experiments~\cite{Werthmuller:2015owc,Witthauer:2017wdb,Witthauer:2017get,Witthauer:2017pcy,Sokhoyan:2018wcd}.
It was explained by {coupled-channel effects} due to $S_{11}(1650)$ and $P_{11}(1710)$ in a $K$-matrix approximation coupled-channel model~\cite{Shklyar:2006xw,Shklyar:2012js}, which was further used to study Compton scattering off {the} proton~\cite{Cao:2017njq}. Alternative interpretations were the {interference} in the $1/2^-$ wave - $S_{11}(1535)$ and $S_{11}(1650)$ - in chiral quark model~\cite{Zhong:2011ti} and Bonn-Gatchina analysis~\cite{Anisovich:2015tla,Anisovich:2017xqg}, and loop contributions from associated strangeness threshold openings~\cite{Doring:2009qr}. The situation would be clarified if the neutron helicity amplitudes of $N^*$ were better constrained. Much progress has been made recently in this direction because of the newly released data of $\gamma n$ reactions~\cite{Strakovsky:2015uji,Briscoe:2019,Anisovich:2017afs}, also $\gamma n\rightarrow K^0\Lambda/K^0 \Sigma^0$ from A2~\cite{Akondi:2018shh}, BGOOD \cite{BGOOD:2021oxp}, and CLAS Collaboration \cite{CLAS:2017gsu}, $\gamma n\rightarrow K^+ \Sigma^-$ from CLAS Collaboration \cite{CLAS:2021hex,CLAS:2020spy}, and $\gamma n \to \pi^0 n$ from A2 Collaboration~\cite{Briscoe:2019cyo},
$\gamma n \to \pi^- p$ from CLAS \cite{CLAS:2017kua} and PIONS@MAX-lab~\cite{Strandberg:2018djk,Briscoe:2020qat} and CLAS Collaboration~\cite{CLAS:2022kta}. So the model analysis of this structure was expected to be refined soon. The signal of its isospin partners was claimed to be present in the $\gamma N \to \eta \pi N$ by Graal~\cite{Kuznetsov:2017xgu}, however, it was not confirmed by CBELSA/TAPS~\cite{CBELSATAPS:2021osa} and A2@MAMI groups~\cite{Werthmueller:2019,Werthmuller:2019teg}. Together with this state, a narrow structure at 1720 MeV close to $\omega p$ threshold was also observed in $\gamma p \to \gamma p$~\cite{Kuznetsov:2015nla}, quasifree $\gamma n \to \eta n$~\cite{Werthmuller:2015owc} and high-precision $\pi p$ elastic data from EPECUR Collaboration~\cite{Alekseev:2014pzu,Gridnev:2016dba}. Whether these states correlate with each other is under investigation. A coupled-channel calculation showed that the narrow structures in Compton scattering $\gamma p \to \gamma p$ are feeble after considering carefully all the contribution of known $N^*$ and $\Delta^*$~\cite{Cao:2017njq}. In this paper as an exemplar of the major obstacles we further demonstrate a full set of beam polarization of proton Compton scattering up to the third resonance region in Sec.~\ref{sec:light}.
Another narrow resonance near {the} $\eta^\prime p$ threshold is possibly {existing} in the GRALL data on the beam asymmetry as discussed by the Bonn-Gatchina approach \cite{Anisovich:2018yoo}.
Considering the connection of $\eta^\prime$ and $\eta$ mesons to gluon dynamics \cite{Bass:2018xmz}, the structures near $\eta^{(\prime)} p$ threshold are of renewed interest.

The tension seems to be relieved when one moves to heavy quark sector.
Since the uplift discovery of the $\chi_{c1}$(3872) by the Belle collaboration~\cite{Choi:2003ue}, a rich spectrum of exotic mesons has brought into an intriguing prospect of hadron physics.
Among them, the charged $Z_{c}$ (or $T_\psi$) states with narrow width play a special role because of their probable $u\bar{d}c\bar{c}$ {composition}.
The $Z_c^+$(4430) with a mass of 4478 $^{+15}_{-18}$ MeV and a width of 181 $\pm$ 31 MeV was found in the $\pi^\pm \psi(2S)$ spectrum of $B^0 \rightarrow K^\mp \pi^\pm \psi(2S)$ by Belle~\cite{Choi:2007wga,Chilikin:2013tch} and confirmed in $\bar{B}^0 \rightarrow K^- \pi^+ \psi(2S)$~\cite{Mizuk:2009da} and $\bar{B}^0 \rightarrow K^- \pi^+ J/\psi$~\cite{Chilikin:2014bkk}.
The LHCb group determined its spin-parity unambiguously to be 1$^+$~\cite{Aaij:2014jqa}. Up to now its other decay modes have been not found.
The $Z_c(3900)$ with a mass of 3887.2 $\pm$ 2.3 MeV and a width of 28.2 $\pm$ 2.6 MeV was discovered in the $J/\psi \pi^{\pm}$ spectrum by BESIII~\cite{Ablikim:2013mio} and Belle~\cite{Liu:2013dau} and confirmed by CLEO-c~\cite{Xiao:2013iha}.
The BESIII collaboration identified it as an isovector state with spin parity $1^+$~\cite{Collaboration:2017njt,Ablikim:2015tbp}, and found its decay channel $D \bar{D}^*$~\cite{Ablikim:2015gda,Ablikim:2013xfr}.
At present, $Z_c$(3900) is the {lowest} $Z_c$ state while $Z_c^+$(4430) is the highest one, limited{, however, only} by the kinematic{al} coverage of {the} accelerator.
The properties of other $Z_c$ states are less known, e.g. the spin-parity of most of them are not well determined~\cite{Cao:2018vmv}.
Moreover, LHCb discovered a much narrower doubly charmed tetraquark candidate $T_{cc}^{+}$(3875) of isoscalar {characteristic} near $D^{*+} D^0 $ threshold, intriguingly stable with respect to the strong interaction \cite{LHCb:2021auc,LHCb:2021vvq}.

The narrow exotic baryons in {the charm sector}, known as pentaquark states $P_c$ (or $P^N_{\psi}$), are predicted within the framework of the coupled channel unitary approach with the local hidden gauge formalism~\cite{Wu:2010jy}. This is confirmed by many models with other prescriptions~\cite{Wu:2010vk,Wang:2011rga,Yang:2011wz}.
The LHCb collaboration {reports} evidently three narrow states $P_c$(4312), $P_c$(4440) and $P_c$(4457) in the $J/\psi p$ invariant mass spectrum of $\Lambda_b^0\to J/\psi p K^-$ decay with the width of $9.8 \pm  2.7 ^{+3.7}_{-4.5}$ MeV, $20.6 \pm  4.9 ^{+8.7}_{-10.1}$ MeV, and $6.4 \pm  2.0 ^{+5.7}_{-1.9}$ MeV, respectively~\cite{Aaij:2015tga,Aaij:2019vzc}.
They also announced {the observation of} another $P_c$(4337) state with a width of 29 $^{+26}_{-12}$ $^{+14}_{-14}$ MeV with a lower significance in $B_s^0 \to J/\psi p \bar{p}$ decays \cite{LHCb:2021chn}.
Their component is likely to be $qqqc\bar{c}$ and their nature is yet under extensive investigation.

Hypernuclei are nuclei within which one or more nucleons are substituted by hyperons, namely $\Lambda$, $\Sigma$ or $\Xi$, which carries a new quantum number, not contained normally inside the nuclei, the strangeness.
The concept of 'hyper' can be extended to exotic mesons, predicted by several framework \cite{Lee:2008uy,Chen:2013wca,Voloshin:2019ilw,Ferretti:2020ewe} as strange exotic mesons, where one quark is replaced by {a} strange quark.
Just recently three $T^\theta_{\psi s1}$ states, namely the $Z_{cs}$(3985) unveiled in $D_s^-D^{*0,+} + D_s^{*-}D^{0,+}$ distribution by BESIII colaboration \cite{BESIII:2020qkh,BESIII:2022qzr} and $Z_{cs}$(4000,4020) found in $B^+ \to J/\psi \phi K^+$ by LHCb group \cite{LHCb:2021uow}, reveal a new dimension to the traditional family of exotic mesons, beyond the hidden-charm and hidden-bottom components.
In the molecular scenario, the $Z_{cs}$(3985) is the ideal candidate of strange partners of the $Z_c$(3900) and $Z_c$(4020) under SU(3)-flavor symmetry \cite{Yang:2020nrt}.
From the global fit to the available data, it is found that present precision is insufficient to disentangle those exotic mesons to be bound or virtual or resonant states.
The existence of $Z_{cs}$(4130), the heavy quark spin symmetry partner of $Z_{cs}$, plays a key role in distinguishing various models and deciphering whether $Z_{cs}$(3985) and $Z_{cs}$(4000) are the same state.
So it is essential to hunt for it at running and future facilities.
Its clue is weakly traced in the data of $\bar{B}_s^0 \to J/\psi  K^- K^+$ at LHCb \cite{Cao:2021ton} and $e^{+}e^{-}\rightarrow K^+ D_{s}^{*-} D^{*0}+c.c.$ at BESIII \cite{BESIII:2022vxd}.
Surprisingly, it would be produced under lower center-of-mass (c.m.) energy of $e^+ e^-$ than that of $Z_{cs}$ due to the inverted coupling hierarchy of triangular singularity at electron-positron annihilation \cite{Cao:2021ton}.
This prediction would be confirmed {by increasing the amount of collected events at BESIII}.
Excitingly, the $X_{0,1}$(2900) (or $T_{cs0,1}$) in the $D^- K^+$ channel were clearly observed by LHCb group as the first charm-strange exotic hadrons with open flavor and without a heavy quark-antiquark
pair \cite{LHCb:2020bls,LHCb:2020pxc}.
The strange pentaquark states $P_{cs}$ (or $P^\Lambda_{\psi s}$) in $J/\psi \Lambda$ spectrum, predicted as $\bar{D}^{(*)} \Xi_c$ molecule and strange partner of $P_c$~\cite{Wu:2010jy}, were also found at LHCb \cite{LHCb:2020jpq,Collaboration:2022boa}.

Molecules {configurations are} the most popular {scenarios for understanding} the nature of these states because of the closeness of corresponding open charm channel, though other explanations are not excluded at all, e.g. tetraquark or hadrocharmonium states, see Refs.~\cite{Chen:2016qju,Guo:2017jvc,Lebed:2016hpi,Esposito:2016noz,Olsen:2017bmm,Liu:2019zoy,Brambilla:2019esw,Guo:2019twa,Chen:2022asf} for review.
A complete spectrum for hadronic molecules seems to be emergent and could be nicely organized by heavy quark spin and flavor symmetries \cite{Guo:2013sya}.
For instance, {a isovector axial-vector multiplet in the} charm and bottom sectors is predicted by further combining SU(3) flavor symmetry for the potential between heavy mesons \cite{Yang:2020nrt,Cao:2020cfx,Meng:2020ihj,Wang:2020htx,Liu:2019tjn,Xiao:2019aya,Peng:2020hql}.

In the bottom sector, two charged states $T^b_{\Upsilon 1}$, namely $Z_b$(10610) and $Z_b$(10650) with the component of $u\bar{d}b\bar{b}$, have been found a decade ago in $\Upsilon(nS) \pi$ (n=1,2,3) and $h_b(mP) \pi$ (m=1,2) spectrum by Belle Collaboration \cite{Belle:2011aa}.
A most recent effort in search of the bottomonium equivalent of the $\chi_{c1}$(3872) state decaying into $\omega \Upsilon$(1S) by Belle II \cite{Belle-II:2022xdi} observed no significant signal for masses between 10.45 and 10.65 GeV.
Several isoscalar states of spin $J=0, 1,2$ with positive parity were also foreseen by different models soon after the discovery of $Z_b$ ~\cite{Guo:2013sya,Cao:2014vma,Dias:2014pva}.
But the strange partners $Z_{bs}$ (or $T_{\Upsilon s}$) and isoscalar analogues have not been found yet.
The double-beauty states as the bottom partner of $T_{cc}^{+}$(3875) was predicted by heavy-quark symmetry based on the observation of $\Xi_{cc}^{++}$ \cite{Karliner:2015ina,Eichten:2017ffp}.
Under SU(3) flavor symmetry for the potential between heavy mesons and baryons, the correspondence of $P_c$ in  {the} beauty sector, labeled as $P_b$ (or $P^N_{\Upsilon}$) here, are supposed to be surely in existence~\cite{Wu:2010rv,Xiao:2013jla,Karliner:2015voa,Karliner:2015ina}.
The analogues in light quark sector, the aforementioned $Z_s$ and $P_s$, are absent experimentally in this jigsaw puzzle,
with the only exception of $\Lambda$(1405) \cite{Jido:2003cb,Mai:2020ltx} and $D_{s0}$(2317) \cite{Guo:2006fu,Liu:2012zya,Chen:2016spr,Yang:2021tvc,Ortega:2021fem,Hao:2022vwt} as a two pole structure of the scattering matrix close to the nominal resonant position.
These facts are challenging our understanding of the strong interaction at low energy, which anticipates a moderate violation of SU(3) flavor symmetry.

Much effort has been taken to the study of pion- and photo-induced reactions for the purpose of searching for these narrow exotic candidates. However, the motivation to study these processes is far beyond this. The photoproduction of these states in two body final states, e.g. near-threshold $J/\psi$ and $\Upsilon$ exclusive photoproduction off the proton, is an exceptional place to exclude their non-resonant possibility. Triangle singularity (TS) is often happened in reactions with three body final states. For instance, various triangle diagrams for $\pi^- p \to J/\psi p \pi^-$~\cite{Liu:2016dli}, $\Lambda_b^0\to J/\psi p K^-$, and $B_s^0 \to J/\psi p \bar{p}$ \cite{Guo:2015umn,Nakamura:2021qvy,Nakamura:2021dix} have been already extensively explored. Though TS can be present in reactions with two body final states, for example the $\gamma p \to V p$ reaction (with $V$ being a vector meson hereafter), it is very hard to satisfy the on-shell conditions required by the TS, as discussed in detail in {the} literature~\cite{Cao:2019kst} and {a} recent review~\cite{Guo:2019twa}. So from the very beginning photoproduction and electroproduction reactions are suggested to disentangle the true resonance nature of exotic states
for the advantage of {being} free of the disturbance of kinematical effects \cite{Wang:2015jsa}.
{If events are statistically abundant, we may move one step further to
nail down the quantum numbers of some of those exotic states
with the help of angular distributions and polarized observables.}

{Moreover, the photo- and electro-production can make maximum use of total energy to search for narrow exotic state of higher mass.
Nearly all exotic states are observed by the electron-positron annihilation and weak decays of $B$, $\Lambda_b$ and their strange partner.
Thus the maximal mass of a charmonium-like or penquark state produced in this way is limited to be around 5.0 GeV by the mass difference between the the ground state bottom-hadrons and $K$ mesons.
The utmost mass of states touched by electron-positron colliders is below 5.0 GeV as well, bounded by the designed c.m. energies.
The photo- and electro-production are likely to extend considerably the mass range up to high excitation region,
another way to probe the internal structure of exotic states.
The clean photo- and electro-production are also complement to prompt production processes at hadron colliders with huge backgrounds and other novel metheods \cite{Ding:2023evu,He:2022rta,He:2021smz}.}

The COMPASS {detector}, abbreviation of COmmon Muon Proton Apparatus for Structure and Spectroscopy, is a fixed-target experiment with muon, pion and proton beams and polarised proton and deuteron targets at the Super Proton Synchrotron (SPS) at CERN.
By using the photoproduction with a muon beam, it covers the range from 7 GeV to 19 GeV in the c.m. energy of the photon-nucleon system.
It has studied not only the aforesaid light exotic mesons, but also the upper limit of $\gamma p \to \chi_{c1} (3872) p$~\cite{Aghasyan:2017utv} and $\gamma p \to Z_c^+(3900) n$~\cite{Adolph:2014hba}.
The GlueX experiment located in Hall D at Jefferson Lab (JLab) measured the $\gamma p\to J/\psi p$ reaction for the first time and set model-dependent upper limits on the branching fractions of $P_c$ decay \cite{Ali:2019lzf}.
The Hall C at JLab proposed to search for the $P_c$ with higher precision by the same reaction \cite{Meziani:2016lhg,Joosten:2018gyo} and published just recently the {data on} differential cross sections \cite{Duran:2022xag}.
The proposed electron-ion collider in {the} US (US-EIC) \cite{Accardi:2012qut,Aschenauer:2014cki,AbdulKhalek:2021gbh} and China (EicC) \cite{CAO:2020EicC,CAO:2020Sci,Anderle:2021wcy} are potential platforms to resolve the nature of those states.
The c.m. energy of {the} latter machine is close to that of COMPASS but with more than one order higher luminosity.

{In this topical review we will first outline the model framework of the electro- and photo-induced reactions in Sec. \ref{sec:framework}.
Afterwards we retrace a hunting for narrow exotic candidates in the light quark sector in Sec. \ref{sec:light}, followed by a detailed inspection of the charm sector in Sec. \ref{sec:charm} and the bottom sector in Sec. \ref{sec:beauty}, respectively.
In particular in last two sections we briefly summarize to-the-date efforts and give an outlook to the future precision frontiers of photo- and electro-production of heavy quarkonium-like states of narrow widths, and stress their impact on the understanding of mystery in light quark sector.
Several critically relevant topics are discussed as well with an emphasis of the model dependence.
We give a short summary and perspective in Sec. \ref{sec:conclusion}.
}

\section{Outline of the Framework }
\label{sec:framework}

\begin{figure}[b]
\centering
\includegraphics[width=8cm,clip]{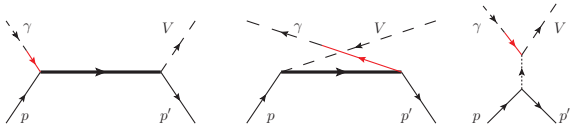}
\caption{The $s$-, $u$- and $t$-channel kernel of $\gamma p\to V p$ with $V$ being $\gamma$, $J/\psi$, $\Upsilon$ or other light mesons or heavy axial-vector mesons. The red lines label $\rho$, $\phi$, $\omega$, $J/\psi$, $\Upsilon$ if VMD model is implemented. The exotic baryon candidates $N^*$, $P_c$ and $P_b$ are potentially produced in $s$-channel $\gamma p\to \gamma p$, $\gamma p\to J/\psi p$ and $\gamma p\to \Upsilon p$, respectively, whereas the exotic axial-vector meson candidates $V$ are likely appearing in $t$-channel.}
\label{fig:channels}
\end{figure}

Since mesons and baryons {are observed as asymptotic QCD} effective degrees of freedom {by} experiment,
the effective models respecting basic symmetries are constructed on the hadron level to describe exclusive photoproduction of meson in $s$-, $u$- and $t$-channel.
The amplitudes of tree diagrams in Fig. \ref{fig:channels} are easily calculated
by means of effective Lagrangian \cite{Penner:2002md,Penner:2002ma}, covariant $L$-$S$ scheme \cite{Zou:2002yy,Zou:2002ar,Cao:2010km}, or {in the} helicity formalism \cite{Albaladejo:2020tzt},
whose details {have been presented extensively in the literature}.
The experimental observables, i.e., the cross sections
and polarization observables, could be easily calculated by partial-wave {techniques} \cite{Ireland:2019uwn}.
In the light quark sector the model parameters are under good control by {a} simultaneous fit to the amount of data of available channels.
The parameters of resonances from low to high excitation are extracted with continuous improvement of accuracy.
Herein we {briefly} outline the essentials of the framework,
{in particular the main feature of technique and reaction kinematics,
but leave the theories and models of hadron structure into other comprehensive reviews ~\cite{Chen:2016qju,Guo:2017jvc,Lebed:2016hpi,Esposito:2016noz,Olsen:2017bmm,Liu:2019zoy,Brambilla:2019esw,Guo:2019twa,Chen:2022asf}.
}.

If a resonant baryon or baryon-like state $R$ is long-lived, 
{the production cross section} in $s$-channel $\gamma p \to R \to V p$ can be simplified to be a Breit-Wigner line shape,
\bea \label{eq:sigmaR}
  \sigma_R  &=& \frac{2\,J+1}{(2\,s_1+1)(2\,s_2+1)} \frac{4\,\pi}{k_{\textrm{in}}^2} \frac{\Gamma^2}{4} \nonumber \\
  && \times \frac{\mathcal{B}(R \to \gamma p) \, \mathcal{B}(R \to V p)}{(W-M)^2+\Gamma^2/4}
\eea
Here $k_{\textrm{in}}$ is the magnitude of three momentum of initial proton in the c.m. frame, $W$ is the c.m. energy of $\gamma p$ system, and {$s_{1,2}$ are the spins} of initial photon and proton, respectively.
This works quite well for the $P_c$ and $P_b$ because of their very large masses in comparison of their widths.
So a product of branching ratios $\mathcal{B}(R \to \gamma p) \, \mathcal{B}(R \to V p)$ can be model independently extracted from the cross section measurements.

The radiative decay width $\mathcal{B}(R \to \gamma p)$ is proportional to the $\mathcal{B}(R \to V p)$ via the vector meson
dominant (VMD) assumption that vector mesons dominate the interactions of hadrons with electromagnetism \cite{Karliner:2015voa,Kubarovsky:2015aaa}:
\be
\mathcal{B}(R \to \gamma p)= \frac{3\,\Gamma(V \to e^+e^-)}{\alpha M_{V}} \sum_L f_L \lf( \frac{k_{\textrm{in}}}{k_{\textrm{out}}} \rg)^{2L+1} \mathcal{B}(R \to V p)  \,
\ee
Here $\alpha$ is the fine structure constant, $L$ is the orbital excitation between the $V$ and the proton, $f_L$ is the fraction of decay in the relative partial wave, and $k_{\textrm{out}}$ is the magnitude of three momentum of final nucleon in the c.m. frame.
Before proceeding further a few remarks shall be made on VMD, which is introduced as an important phenomenological concept before the era of quantum chromodynamics \cite{Sakurai:1960ju}.
The model has been validated for the lightest vector mesons $\rho$, $\omega$ and $\phi$ as the constitution of the hadronic components of the physical photon \cite{Meissner:1987ge,Leupold:2012qn},
though it is eliminated as a possible description of deep inelastic scattering \cite{Friedman:1991nq}.
In {the} light quark regime, VMD is not {a} prerequisite for the framework since electromagnetic helicity amplitudes could be fixed with controlled uncertainties by plenty of photoproduction data.
Generalization of the {VMD model to heavy vector quarkonium has been proposed to drastically fail} \cite{Xu:2021mju}. 
To identify the cases where VMD fails is of its own importance. A future effort would improve the estimating the momentum dependent photon-to-quarkonium transition strength. 

The exotic meson candidates $V$ can be produced through $t$-channel of $\gamma p\to V p$
and then reconstructed and analyzed by their subsequent decay.
At high energies they are searched for by diffractive process as mentioned for {the} $a_1$- and $\pi_1$-meson.
The mechanism for the production of axial-vector Charmonium states is further detailed below in Sec. \ref{sec:charm}.
As the non-resonant background {to the $s$-channel contributions}, the $t$-dependence {contributions} (or angular distributions) can be model-dependently calculated.
The total cross section {due to the $t$-channel processes is appropriately estimated by}
\be \label{eq:sigmaV}
  \sigma_V = \mathcal{N} W^{\delta(Q^2)} = \mathcal{N} W^{\alpha+\beta \ln (Q^2+M_V^2)}
\ee
which is suggested by the empirical formula of deeply virtual meson production (DVMP) $\gamma^* p \to V p$~\cite{Favart:2015umi}. Here the units of $M_V$ and $W$ are in GeV and that of the photon virtuality $Q^2$ in GeV$^2$.
The merit of this simple parameterization is that it is applicable to various mesons with proper $Q^2$ dependence. The parameters $\alpha$ and $\beta$ have been determined by the DVMP data to be $\alpha = 0.31 \pm 0.02$ and $\beta = 0.13 \pm 0.01$~\cite{Favart:2015umi}.
The corresponding $\delta(Q^2=0) = 0.89$ is confronted with the perturbative QCD prediction $\delta \sim 1.7$~\cite{Frankfurt:1998yf}, {whose difference is not satisfactorily explained.}
The normalization $\mathcal{N}$ of $\gamma p \to \Upsilon p$ is determined by the data at high energies to be $2.62 \pm 0.38$ {fb}, where the experimental error of $W$ is not {included} \cite{Cao:2019gqo}. Its extrapolation to low energies {is} suggested to be an upper limit of the production rates \cite{Cao:2019gqo}.
A more appropriate evaluation for the near-threshold region is to include the two-body phase space factors for Eq. (\ref{eq:sigmaV}), or use the formula based upon Pomeron exchange \cite{Levin:1998pk} with the intercept of Regge trajectory close to 1 \cite{Donnachie:1984xq,Donnachie:1992ny},
or alternatively two-gluon and three-gluon exchange \cite{Brodsky:2000zc}.
These schemes are widely applied for $J/\psi$ {of which} the normalization $\mathcal{N}$ is well fixed by {the} data from near-threshold up to $W =$ 100 GeV \cite{Meziani:2016lhg}.

\begin{figure}[t]
\centering
\includegraphics[width=8cm,clip]{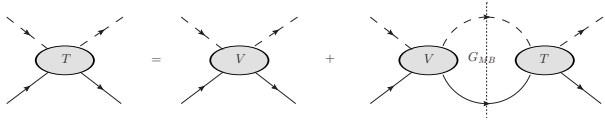}
\caption{A diagrammatic representation of $K$-matrix approximation coupled-channel model. Dashed lines are mesons or photon, and solid lines are baryons.}
\label{fig:ccfeymann}       
\end{figure}

{At} the low-energies several hidden and open strange channels, e.g. $\gamma N$, $\pi N$, $2\pi N$, $\eta N$, $\omega N$, $\pi N$, $K \Lambda$, and $K \Sigma$, are involved and the available data {are} sufficient for a more comprehensive analysis. A coupled-channel unitary Lagrangian model could be constructed with an input of tree level diagrams $V_{fa}$ in Fig,~\ref{fig:channels},
\be \label{eq:cc}
T_{fi}=V_{fi}+V_{fa}G_{ab}T_{bi}
\ee
as {graphically} illustrated in Fig.~\ref{fig:ccfeymann}. {The $i$, $f$ and $a(b)$ are the initial, final, and intermediate states, respectively.}
Besides partial wave {analyses} on the basis of proper {parameterization of amplitudes} \cite{Workman:2012hx,Anisovich:2017bsk,Anisovich:2015tla,Anisovich:2017xqg}, 
{several} dynamical frameworks are proposed for the interaction kernel $V_{fa}$ in order to resolve approximately the coupled-channel equation \cite{Huang:2011as,Ronchen:2022hqk,Wang:2022osj,Kamano:2013iva,Kamano:2016bgm,Kamano:2019gtm}.
Alternatively, another $K_{fa}$ kernel could be defined as,
\be
K_{fi} = V_{fi} + {V_{fa}\textrm{Re}G_{ab}K_{bi}}
\ee
As a practical approach, the $K$-matrix approximation assumes that the real part of
the propagator $G_{ab}$ is vanishing and the coupled-channel equation is accordingly reduced to \cite{Penner:2002ma,Penner:2002md}:
\be
 T_{fi}=K_{fi}+iK_{fa}\textrm{Im}G_{ab}T_{bi}
\ee
in terms of $K_{fa} = V_{fa}$ accommodating the reaction mechanism of the pion- and photo-induced reactions in the resonance region \cite{Shklyar:2004ba,Shklyar:2006xw,Shklyar:2012js,Shklyar:2014kra}.
The summation over intermediate states $a(b)$ is running only over hadronic states but neglecting the $\gamma N$ so gauge invariance of the Compton amplitude is easily retained \cite{Penner:2002md,Cao:2017njq}.
Thus, unitarity holds with {the merit of} technical simplicity and flexibility but at the cost of analyticity.
As a result, all states are treated as Breit-Wigner resonances as {the same} in Eq.~(\ref{eq:sigmaR}) and neither of them is dynamically generated.
The framework is readily applicable to high energies
if several channels, e.g. various hidden and open charm or bottom final states are involved into the analysis.
{Especially, the kernel is envisaged to incorporate the dynamical model of internal structure.}
However, {by the} lack of data, the analysis at present {is restricted to the single channel case}
with the aim to estimate the photoproduction rates of various quarkonium and exotic quarkoniumlike states in the kinematic{al} regime of running fixed-target machine and future electron-ion colliders.
A few states of particular interest are used as benchmarks with the help of the limited experimental information.
{The reaction dynamics, sensitive to specific process, are separately discussed in later sections.}

With the input of $t$-dependent photoproduction cross sections,
the differential cross section of exclusive meson electroproduction can be directly calculated under {the} one photon approximation~\cite{Joosten:2018gyo,Meziani:2016lhg}:
\be
\frac{\textrm{d} \sigma_{ep\to eVp}}{\textrm{d} Q^2 \textrm{d} y \textrm{d} t} = \Gamma_\textrm{T} (1+\epsilon\, R_\textrm{L}) f(Q^2) \frac{\textrm{d} \sigma_{\gamma p\to Vp}}{\textrm{d} t}
\ee
where $y$ is the fractional energy of the incoming electron
transferred to the virtual photon in the target rest frame.
The parameterization of logitudinal-to-transverse ratio $R_\textrm{L}$ and form factor $f(Q^2)$ are widely studied in the literatures~\cite{Martynov:2002ez}.
Generally the $R_\textrm{L}+1(m < 0)$ and $f(Q^2) (m > 0)$ are written as,
\be \label{eq:VMDformfactor}
\left( \frac{n \, M_V^2}{n \, M_V^2 + Q^2} \right)^m
\ee
{where the} parameters can be determined by a global fit to vector meson electroproduction {data}. The study of available $e p \to e J/\psi p$ data gives valuable information on the $Q^2$ dependence of heavy quarkonium production~\cite{Martynov:2001tn,Martynov:2002ez}. This is widely used by the simulation of JLab-12 \cite{Arrington:2021alx} and US-EIC~\cite{Joosten:2018gyo,Meziani:2016lhg}.
The virtual photon flux is defined as,
\be \label{eq:flux}
\Gamma_\textrm{T} = \frac{\alpha_{em}}{2\,\pi} \frac{y^2}{1-\epsilon} \frac{1-x}{x Q^2}
\ee
where the polarization is $\epsilon = (1-y-\frac{1}{4}\,\gamma^2 y^2)/(1-y+\frac{1}{2} y^2 +\frac{1}{4} \gamma^2 y^2 )$ with  $\gamma = {2xM_N}/{Q}$.
For high energies $\epsilon$ is approaching unity, for instance the averaging polarization $<\epsilon> \simeq 0.99$ at HERA.
Other variables are:
\be
x = \frac{Q^2}{W^2-M_N^2+Q^2} \,, \qquad y = \frac{Q^2}{x(s-M_N^2)}  
\ee
The kinematic{al} coverage is limited to:
\bea
1 \geq &y& \geq \frac{Q^2+(M_V+M_N)^2-M_N^2}{s-M_N^2} \nonumber \\
\sqrt{s-Q^2} \geq &W& \geq {M_V+M_N} \nonumber \\
t_0(y,Q^2) \geq &t& \geq t_1(y,Q^2) \nonumber \\ 
\frac{m_e^2 y^2}{1-y} \leq &Q^2& \leq ys + (1-y) M^2 - (M_V+M_N)^2 \qquad \qquad \nonumber 
\eea
with the definition of
\bea
t_{0,1} &=& \frac{(Q^2+M_N^2)(M_V^2-M_N^2)\pm \lambda(Q^2) \lambda(M_V^2)}{2 W^2} \nonumber \\ && - \frac{1}{2}(W^2+Q^2-M_V^2-2 M_N^2)
\eea
where $\pm$ corresponds to 0 and 1, respectively. The K\"{a}ll\'{e}n triangle function {are} $\lambda^2(x,y,z) = x^2 + y^2 + z^2 -2xy - 2yz -2zx$,
$\lambda(Q^2) = \lambda(W^2,-Q^2,M_N^2)$, and $\lambda(M_V^2) = \lambda(W^2,M_V^2,M_N^2)$.
As a substitute the $t$-integrated photoproduction cross section can be used~\cite{Klein:2019avl}:
\be
\sigma_{e p \rightarrow e V p} = \int \frac{\textrm{d}W}{W} \int \textrm{d}k \int \textrm{d}Q^2 {\mskip 2.5mu} \Gamma(k,Q^2) {\mskip 2.5mu} {f(Q^2)} {\mskip 2.5mu} { \sigma_{\gamma p\rightarrow V p} (W)}
\ee
where $\Gamma(k,Q^2)$ is flux under equivalent photon approximation with $k$ being the photon energy.

\section{Exotic candidates in light quark sector }
\label{sec:light}

\begin{figure*}
\centering
\includegraphics[width=16cm,clip]{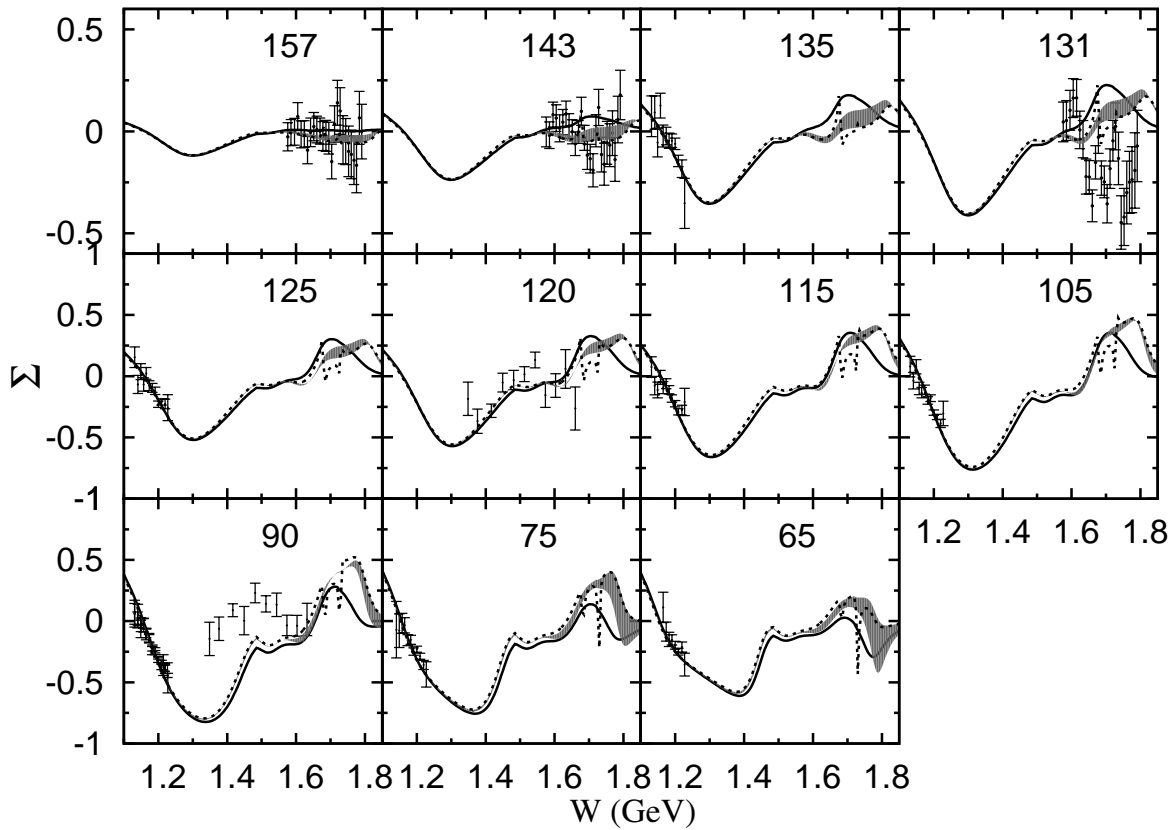}
\caption{The beam polarization of proton Compton scattering versus c.m. energies W for different angular bins (in unit of degree as labeled in each figure). Solid lines are the result with the parameters in Ref.~\cite{Cao:2013psa}, and shaded area are the improved result with adjusting the helicity couplings of $D_{33}$(1700) and $F_{35}$(1905) resonances. Dotted lines are the one for adding $S_{11}$(1680) and $P_{11}$(1720). The data in the scattering angles $131^{\circ}$, $143^{\circ}$, and $157^{\circ}$ are from GRAAL collaboration~\cite{Kuznetsov:2015nla} and others are referred to the compilation in Ref.~\cite{Penner:2002md}.}
\label{fig:ggpsigall}       
\end{figure*}

\begin{table}[b]
\centering
\caption{The parameters of two exotic $N^*$ added in the coupled-channel model. The Breit-Wigner (BW) masses and total widthes $\Gamma_{tot}$ are given in MeV. The sign of electromagnetic helicity amplitudes is not determined.}
\label{tab:Nstar}       
\begin{tabular}{l|l|l|l}
 \hline
 $N^*$               & BW mass & $\Gamma_{tot}$ & $A^p_{\frac{1}{2}}$ ( $10^{-3}$ GeV$^{-1/2}$) \\
 \hline
   $S_{11}$(1680)    & 1681    & 2 $\pm$ 1      & 32 $\pm$ 10  \\ 
 \hline
   $P_{11}$(1720)    & 1726    & 2 $\pm$ 1      & 35 $\pm$ 10  \\ 
 \hline
\end{tabular}
\end{table}

In the light quark sector {as probed in the $\pi N$ and $\gamma N$ reactions} the resonant line shape and non-resonant background are much more involved than those parameterizations in Eqs. (\ref{eq:sigmaR}) and (\ref{eq:sigmaV}).
Before attempting to search for narrow exotic baryon resonances amongst {a considerable amount of the} data, one has to exclude other possibilities, e.g. interference, threshold openings or triangle diagrams, in a sophisticated but reliable manner.
Afore introduced coupled-channel model in K-matrix approximation is well suited for this purpose.
It excavates the known resonances by fitting both isospin $I = 1/2$ and $I = 3/2$ partial waves to the available data.
Other dynamical approaches respecting analyticity are successfully constructed, namely the J\"ulich-Bonn-Washington model \cite{Ronchen:2012eg,Ronchen:2018ury}
and the ANL-Osaka model\cite{Kamano:2013iva,Kamano:2019gtm}.
The former has already extended to study the electroproduction of the $\pi$ and $\eta$ mesons \cite{Mai:2021vsw,Mai:2021aui} and just recently $K \Sigma$ photoproduction \cite{Ronchen:2022hqk}.
The latter is also used to extract $\Lambda^*$ and $\Sigma^*$ from $K^- p$ reactions \cite{Kamano:2014zba,Kamano:2015hxa}, whose result together with Bonn-Gatchina solution \cite{Matveev:2019igl,Sarantsev:2019xxm,Anisovich:2020lec} are encountering {the rather complex structure of the $\Lambda$(1405) state}.
Incidentally, conventional wide resonances are incorporated in all models, though the spectrum would differ case by case.
As also noted, strong evidence is claimed for new wide resonances near 1900\,MeV in $\gamma p \to K \Lambda$ by the Bonn-Gatchina {group} \cite{Anisovich:2017bsk,Anisovich:2017ygb}.
This relieves the famous shortcoming of \textit{missing resonances} in the conventional quark model,
which predicted more baryonic states of three-quark than seen in the $\pi N$ and $\gamma N$ scattering.
{A} similar issue reappears in lattice QCD~\cite{Edwards:2011jj} and Dyson-Schwinger calculations~\cite{Eichmann:2016hgl,Chen:2017pse} under the situation of unphysical $\pi$ mass.
Contrarily, quark-diquark models give rise to too less states, for instance, failing to accommodate  {$P_{13}(1900)$} and {$F_{15}(2000)$} states, though remedy would be put forward.

Besides the narrow structures in several channels of pseudo-scalar meson as outlined in Sec. \ref{sec:intro},
those signals in Compton scattering off the proton in the resonance region need to be carefully analyzed.
As a consequence of smallness of the electromagnetic couplings constant, the electromagnetic reactions {decouple} essentially from the hadronic ones in a coupled-channel model.
This is realized for the first time with the help of the techniques at hand after fixing properly the isospin $I = 3/2$ amplitude by the $K \Sigma$ channel~\cite{Cao:2013psa}.
It results into a refined extraction of the amplitudes of Compton scattering off the proton \cite{Cao:2017njq}.
After a full combined analysis of pion-and photo-induced reactions, beam polarization of proton Compton scattering  can be fairly described without free parameters as shown by the solid lines in Fig. \ref{fig:ggpsigall}.
Selected angular bins have been already published \cite{Cao:2017njq}.
The leading contribution stems from $P_{33}$(1232) and $D_{13}$(1520) in first and second resonance region.
As shown by the shaded area, the agreement is systematically improved if
adjusting a bit the helicity couplings of $D_{33}$(1700) and $F_{35}$(1905) resonances,
which is  predominant in the energy range between 1.6 GeV and 1.8 GeV.

Therefore the {room left} to host new states is lower than 1$\sigma$ {if judged by the} statistical significance.
Still, two exotic $N^*$ {states seen} closely above the $K\Lambda$ and $\omega N$ thresholds are added into the analysis as {indicated} by the dotted lines in Fig. \ref{fig:ggpsigall}.
Their masses are fixed {by comparison to corresponding structures observed in $\eta p$} channels and $\pi p$ elastic differential cross sections.
Other parameters are extracted in Table~\ref{tab:Nstar}, of which the errors are mainly driven by the data of the scattering angles $131^{\circ}$, $143^{\circ}$, and $157^{\circ}$ from GRAAL collaboration~\cite{Kuznetsov:2015nla}.
Their electromagnetic helicity amplitudes in Tab. \ref{tab:Nstar} are found to be of moderate magnitude.

Apart from {the} resonance region,
the nucleon Compton scattering at low energies is a probe of the nucleon's polarizabilities, a measure of their response to an external
electromagnetic field of moderate magnitude \cite{Hagelstein:2015egb}.
Similarly a partial-wave analysis of {the world data set} below the pion-production threshold is accomplished \cite{Krupina:2017pgr}.
The $\Delta$(1232) contribution to the scalar and spin polarizabilites is noticeable, and the $D_{13}$(1520) plays a role {in} the proton's magnetic polarizability~\cite{Eichmann:2018ytt}.
The effect of {any} other resonances is invisible.
As a result, {full understanding of nucleon resonances is established} in Compton scattering from low energies where {chiral perturbation} theory is applicable, up to 1.8 GeV in which a coupled-channel effective Lagrangian model shall be constructed.
Our {K-matrix analysis of the resonance region gives insight into the} highly non-trivial structures in Compton scattering, however, calling for higher precision data.
{In this classic case the obstacles and complexities have been demonstrated which are encountered in the hunting for} narrow exotic states in the light quark sector.
In particular, it reflects the insufficiency of our understanding of the
baryons as relativistic three-quark bound states~\cite{Eichmann:2016yit},
in which {other effects like} diquark correlations \cite{Barabanov:2020jvn}, {gluonic admixtures and vacuum polarization} are also important.

In a word, after a lot of effort none of the light narrow exotic states is firmly established.
It is foreseen that {once high statistics data will become} available in future, partial wave {analyses will be} feasible for
diffractive processes at COMPASS and CLAS \cite{JPAC:2021rxu}
and annihilation reactions at BESIII \cite{BESIII:2020nme} and at PANDA \cite{PANDA:2009yku}.
Other reactions and decays to study them {are} suggested from the theoretical perspective~\cite{Cao:2014mea,Lebed:2015dca,He:2017aps,An:2018vmk,Gao:2017hya}.
A search in the electroproduction of the nonstrange channels is viable when more measurements {will become} available \cite{Mai:2021vsw,Mai:2021aui,Blin:2021twt}.
Whether they are accessible at EicC and US-EIC needs careful investigation by means of available tools.

\section{Exotic candidates in Charm sector: } 
\label{sec:charm}

\begin{figure}[htb]
\centering
\includegraphics*[width=0.5\textwidth]{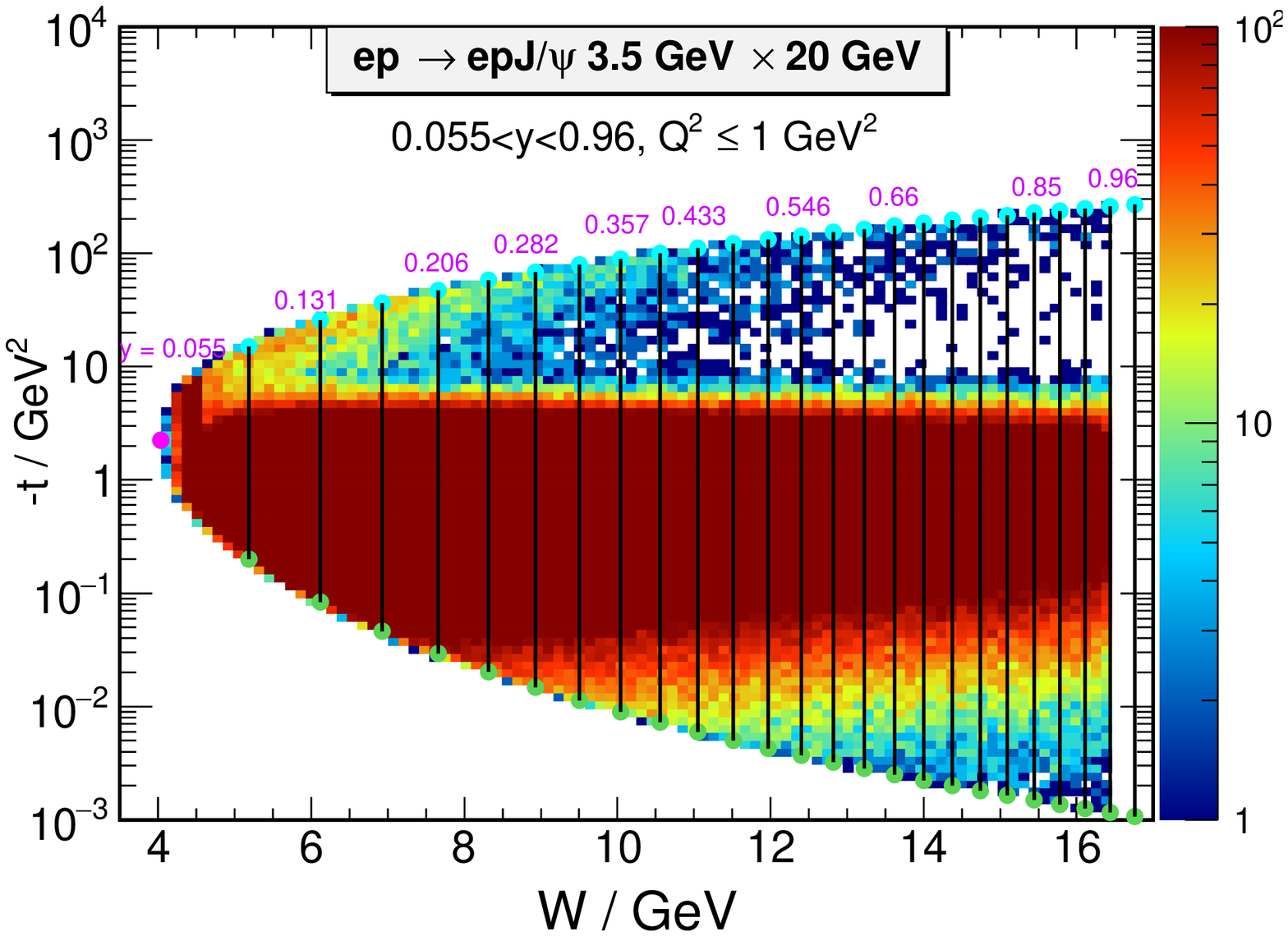}
\includegraphics*[width=0.5\textwidth]{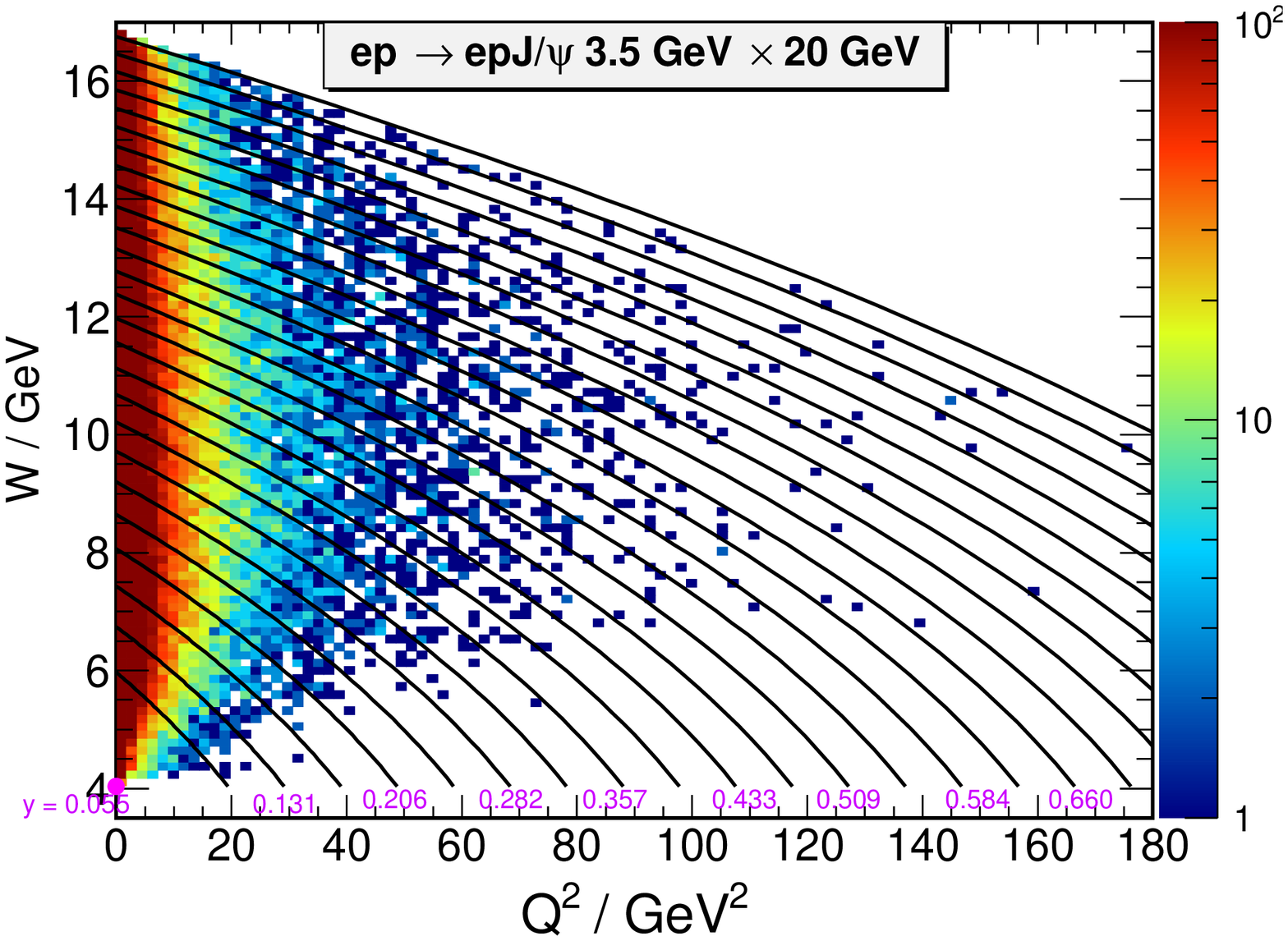}
\caption{Events of $e p \to e J/\psi p$ in $t$-$W$ (Top) and $W$-$Q^2$ (bottom) plane under kinematical coverage of EicC. The signals of $P_c$(4312), $P_c$(4440) and $P_c$(4457) are included in the simulation.}
\label{fig.Jpsi}       
\end{figure}

In the heavy quark sector the {resonance} line shape and non-resonant background are much simpler
because of the narrow width of the established states.
Yet, coupled-channel analysis is desirable with {the} aim to understand the nature of states {by exploring } their pole structures.
The effort along this direction is made in several cases like $P_c$ \cite{Du:2021fmf}, $\chi_{c1}$(3872) \cite{Kalashnikova:2005ui,Ortega:2009hj,Ferretti:2018tco}, $Z_c$(3900) \cite{Du:2022jjv} and $T_{cc}^+$ \cite{Albaladejo:2021vln,Du:2021zzh}.
A complete analysis is postponed in most situation by the shortage of data,
which are usually available only for the dominant channel.
The single channel framework in Sec. \ref{sec:framework} is straightforwardly applied to the electroproduction of these states with the input of photoproduction cross sections if experimentally available.
Otherwise the radiative decay width $\mathcal{B}(R \to V p)$ is required to be theoretically estimated, however, definitely of big model dependence.

The pentaquark states are unveiled via their decay to $J/\psi$ plus proton or hyperon.
Even before and also soon after the discovery of $P_c$
it is speculated that $P_c$ can be photoproduced {in the $s$-channel reactions as} in Fig. \ref{fig:channels} with the
photon converted first to $J/\psi$ \cite{Huang:2013mua,Wang:2015jsa,Kubarovsky:2015aaa,HillerBlin:2016odx}.
The conventional reconstruction route would be hidden charm channels involving $J/\psi$ or $\eta_c$ \cite{Sakai:2019qph}.
Considering that they would decay more favorably into open charm channel, the reaction $\gamma p\to \bar{D}^0 \Lambda^+_c$ {is} investigated
 within an effective Lagrangian approach \cite{Huang:2016tcr} and the Regge-plus-resonance model \cite{Skoupil:2020tge}.
The non-resonant background can be parameterized as the $t$-channel diagram with gluon, Pomeron or meson exchange, {respectively,}
whose angular distributions are different from those of the pentaquark states.
It is naturally conjectured that the differential cross sections fixed by {the} helicity dependence can be used to disentangle the spin and parity of these resonances \cite{Wang:2015jsa}.
In light of these studies, the
electroproduction of pentaquark states in electron-proton collision is explored \cite{Joosten:2018gyo,Yang:2020eye,Xie:2020niw}.
Hopefully the signal of pentaquark would be increased under a proper kinematic{al} cut \cite{Yang:2020eye}, as {examplified by the results obtained with the} lAger Monte Carlo package \cite{Joosten:2018gyo,Joosten:lager} in Fig.~\ref{fig.Jpsi}.
However the {magnitude of the production cross section} is severely dependent on the VMD assumption and form factors of the interaction vertices,
though the computed total cross sections seems to be sizeable.
Fortunately available data have already imposed valuable constraint on the properties of $P_c$ states \cite{Burns:2021jlu}.
More specifically, the JLab photoproduction data {display} no sizable peaks upon the non-resonant $t$-channel process in the cross section \cite{Ali:2019lzf,Duran:2022xag,Arrington:2021alx}.
This puts {an} upper bound of photoproduction cross sections of $P_c$ or model-dependent $\mathcal{B}(P^+_c \to J/\psi p)$,
which are proceeding toward {a size of} several nb or a few percentage, respectively.
On the other hand, as derived by LHCb fit fractions, e.g. for $P_{c}$(4312) and $P_{cs}$(4338) \cite{Aaij:2019vzc,Collaboration:2022boa},
\bea \nonumber \label{lowerlimit}
  \mathcal{B} (\Lambda^0_b \to P_c^+ K^-) \mathcal{B}(P^+_c \to J/\psi p) = 0.96^{+1.13}_{-0.39} \times 10^{-6} &&  \\ \nonumber
  \mathcal{B} (B^- \to P_{cs}^0 \bar{p}) \mathcal{B} (P_{cs}^0 \to J/\psi \Lambda) =  1.83 \pm {0.33} \times 10^{-6}  &&
\eea
if using PDG values of $\mathcal{B} (\Lambda^0_b \to J/\psi p K^-)$ and $\mathcal{B} (B^- \to J/\psi \Lambda \bar{p})$, respectively \cite{ParticleDataGroup:2022pth}.
So the upper limit on weak decay $\mathcal{B} (\Lambda^0_b \to P_c^+ K^-)$ (or $\mathcal{B} (B^- \to P_{cs}^0 \bar{p})$) \footnote{Noted that the decays of $B^- \to P_{cs}^0 \bar{p}$ and $\Lambda^0_b \to P_c^+ K^-$ are related through an approximate dynamical supersymmetry between the anti-quark ($\bar{u}$) and the diquark ($ud$) \cite{Amano:2021spn}.}
implies valuable lower limits on $\mathcal{B}(P^+_c \to J/\psi p)$ (or $\mathcal{B} (P_{cs}^0 \to J/\psi \Lambda))$, surprisingly in the same level of $ 0.5\% \sim 0.05\%$ \cite{Cao:2019kst}.
The ${\gamma}p\rightarrow K^+ P_{cs}(4459)$ occurring through $t$- and $u$-channel {is} estimated to be of the order of several pb \cite{Cheng:2021gca}, one to two orders of magnitude lower than that of the $P_c$.

The electroproduction at electron-ion colliders will push forward this precision frontier into the lower bounds.
The cross section is roughly two orders of magnitude smaller than that of direct photoproduction
for the sake of the electromagnetic coupling in Eq. (\ref{eq:flux}),
but it is compensated by the high luminosity of facility.
Around two million $J/\psi$ exclusive events below $W =$ 20 GeV can be reconstructed under the integrated luminosity of 50 fb$^{-1}$, one percentage among which at most are possibly from the $P_c$'s decay.
Around 90\% events are accumulated in the range of $Q^2 < $1.0 GeV$^2$ due to the $Q^{-2}$ suppression in photon flux, {confirmed by the simulation of eSTARlight generator \cite{Li:2022vjx}}.
In order to measure this quasi-real region with excellent acceptance and
reconstruction efficiency, the detector will be designed
with solid angle coverage, outstanding hadron identification in the forward angle and vertex detector for decay topology \cite{Bylinkin:2022rxd,Adkins:2022jfp}.
The precision close to the threshold is the key issue,
and a cutting edge design of the interaction region would unveil how low $Q^2$ and $W$ (or $y$) could be achieved.
The {kinematically} allowed minimum $Q^2$ is around 10$^{-7}$ GeV$^2$,
with a potential reach of 10$^{-5}$ GeV$^2$ by {a dedicated design of future facilities.} 
The lower energy of the facility leads to more central production of the midrapidity
with a potential sacrifice of the yield \cite{Anderle:2021wcy,CAO:2020Sci}, see Fig.~\ref{fig.Jpsi}.
The measurements of the very close-to-threshold region will be restricted by detection of particles decaying at nearly rest.

After accounting for the branching ratios ($\mathcal{B}(e^+ e^-) \simeq 6\%$) and detection efficiency ($30\% \sim 60\%$),
the yielded events of $J/\psi$ and higher charmonium in exclusive photoproduction are adequate for an accurate study
with a sensitivity below {1 pb}.
{Thus, the collider} mode with polarized electron or proton beam will hopefully
disclose spin {and} parity of $P_c$ \cite{Winney:2019edt}, given that {these states are indeed exists as the genuine ones}.
Their electromagnetic transition form factor would effectively {reveal the} inner structure
through the utilization of the high $Q^2$ events up to 10 GeV$^2$.
Furthermore, the statistics in the open charm channel and $\eta_c$ channel are believed to be even more {illuminating},
{hence calling} for more study on the {reconstruction} efficiency of $\eta_c$, $\bar{D}^0$ and $\Lambda$ by their weak decays.
Semi-inclusive electroproduction is another alternative to collect more events \cite{Yang:2021jof,Winney:2022tky},
especially useful for those states of lower rates, e.g. double-charm states $T_{cc}$ \cite{Shi:2022ipx}.

Several exotic meson candidates are established via their decay to Charmonium ($J/\psi$ or $\psi(2S)$) plus another light meson $V^\prime$.
It is natural to assume that they can be photoproduced through $t$-channel $V^\prime$-meson exchange in Fig. \ref{fig:channels} with the photon converted first to Charmonium,
{of which the} coupling can be calculated by the dilepton decay width of the Charmonium.
Their decays to two light mesons or radiative decay to a light meson are calculated to be not small, however, {awaiting experimental confirmation.} 
For instance {either} significant signals are found for $Z_c(3900)^\pm\to\omega\pi^\pm$ by BESIII \cite{BESIII:2015wge}, {or} $\chi_{c1}(3872) \to \phi\phi$ by LHCb \cite{LHCb:2017ymo},{or} $\pi^+ \pi^- \pi^0$ by Belle \cite{Belle:2022puc}.
This fact casts doubt on {the applicability of} VMD.
Anyway, following closely the information from experimental side, the $\gamma p\rightarrow \chi_{c1}(3872) p$ can be proceeded by $t$-channel vector mesons exchange (e.g. $\rho$, $\omega$ and $J/\psi$ \textit{et al.})~\cite{Albaladejo:2020tzt}.
A {similar} mechanism is appropriate {for} several $X$ states in $J/\psi \phi$ spectrum found by LHCb~\cite{Aaij:2020tzn,LHCb:2021uow}.
The $\gamma p\rightarrow Z_c^+(3900) n$ and $\gamma p\rightarrow Z_c^+(4430) n$ can be populated by $t$-channel Regge exchange~\cite{Galata:2011bi} or charged mesons (e.g. $\pi^+$ and $a_0^+$ \textit{et al.})~\cite{Lin:2013mka,Liu:2008qx} in the phenomenological approaches.
The maximal cross section of those non-strange states reconstructed by the established channel is below 0.1 nb, see Fig.~\ref{fig:photoZc3900} in Appendix \ref{apdx:model} for $Z_c^+(3900)$ detected by the $J/\psi \pi^+$ decay.
The $\gamma p\rightarrow Z^+_{cs} \Lambda$ can be produced by $t$-channel $K$-meson exchange \cite{Cao:2020cfx},
whose maximal cross section is around one to two orders of magnitude lower than that of the $Z_c^+(3900)$ state.
The background contribution is mainly from $t$-channel Pomeron exchange and thought to be small in the {kinematical region of the signal}, usually around 1 GeV above production threshold.
In view of these photoproduction calculations, the electroproduction of $Z_c(4430)$ is simulated by eSTARlight Monte Carlo generator~\cite{Klein:2019avl}.
The spread of events within phase space is analogous to those non-resonant $J/\psi$ production in Fig.~\ref{fig.Jpsi} because of the identical $t$-channel nature.
Since the spin-parity {of} $Z_c^+(3900)$ and $\chi_{c1}(3872)$ have been {encircled} by BESIII and LHCb, their electroproduction {is} in fact of little model dependence with the input of photoproduction cross sections from COMPASS \cite{CAO:2020EicC}, see appendix \ref{apdx:model} for the details.
These $Z_c$ and $Z_{cs}$ states can be reconstructed by their open charm decay, e.g. $D_{(s)} \bar{D}^{(*)} + c.c. $  as well.
The exclusive cross section of double-charm states through central diffractive process in the ${\gamma}p \rightarrow D^+ T_{cc}^- {\Lambda}_c^+$ reaction is around 1 pb \cite{Huang:2021urd}, calling for higher luminosity of electron-ion colliders.
An additional meson needs to {be detected to the expense of} a reduction of {the} overall efficiency in comparison to search for $P_c$, for instance $\gamma p\to {D}^{*} \bar{D}^0 \Lambda^+_c$ or $\pi^0 J/\psi p$.

The theoretical framework is far from being completed for photo- and electro-production at high energies.
While the {single} channel analysis is not unrealistic in Fig. \ref{fig:ccfeymann}, a more complete approach is proposed to include coupled-channel effect.
{Unitary} and analyticity are easily {maintained in} the non-relativistic approximation {when the lowest channel is opening}.
Even {in the} very close-to-threshold regime, the coupled-channel effect would be essential for explaining possible structures \cite{Xiao:2015fia,Du:2020bqj},
though the separation is at least 110 MeV between neighboring channels, e.g. the $\eta_c p$, $\psi$(2S)$p$, $\bar{D} \Lambda_c$, $\bar{D}^* \Lambda_c$, $\bar{D} \Sigma_c$ and $\bar{D}^* \Sigma_c$.
The angular distributions of final meson and proton are dependent on the assigned spin-parity $J^P$.
The coupled-channel partial wave formalism renders the feasibility of discriminating multiplets of different $J^P$ as long as enough statistics are gathered.
Besides the criticism of VMD in general \cite{Xu:2021mju},
whether photon is transformed mainly to charmonium or light vector meson is questioned at high energies \cite{Wu:2012wta,Wu:2019adv}.
To say the least, the precise data {expected from} the future collider would for the first time discern these mechanism and examine the fidelity of VMD  for heavy vector-mesons.
The unambiguous goal is to scrutinize the model independent lower limits on $\mathcal{B}(P^+_c \to J/\psi p)$
and its conformity with electroproduction wihin VMD.
Luckily, these model uncertainties do not affect much the estimation of electroproduction cross section
which is weakly dependent on $Q^2$ for heavy quarkonium.

\section{Exotic candidates in Beauty sector: } 
\label{sec:beauty}

\begin{figure}[b]
\centering
\includegraphics*[width=0.5\textwidth]{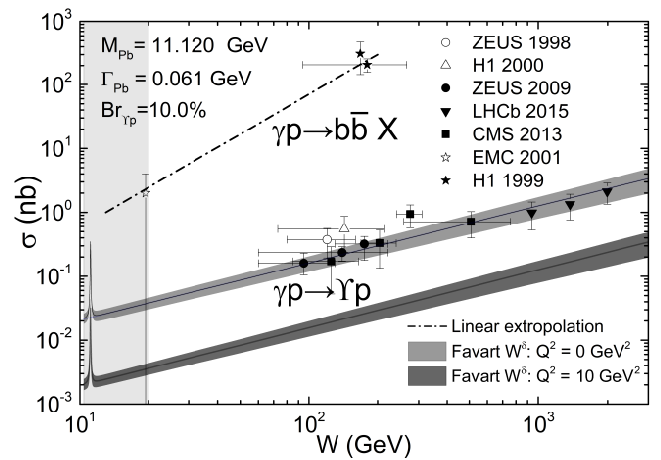}
\caption{The cross section of $\gamma p \to b\bar{b}X$ and $\gamma p \to \Upsilon p$ as a function of $\gamma^* p$ energies $W$.
The pale shaded area is the EicC energy region.
The error band is from the uncertainties of three parameters in Eq. (\ref{eq:sigmaV}) from Favart \textit{et al.}~\cite{Favart:2015umi} without considering those of $W$ and $P_b$~\cite{Wu:2010rv,Karliner:2015voa}.
The $\gamma p \to \Upsilon p$ data are from LHCb (solid inverse triangle~\cite{Aaij:2015kea}), ZEUS (open circle~\cite{Breitweg:1998ki}, solid circle~\cite{Chekanov:2009zz}), H1 (open triangle~\cite{Adloff:2000vm}), CMS(solid square~\cite{CMS:2016nct}).
The $\gamma p \to b\bar{b}X$ data are from EMC (open star) \cite{EuropeanMuon:1981hfd} and H1 (solid star)~\cite{Luders:2001nb,Adloff:1999nr}.
}
\label{fig:addinUpsilon}
\end{figure}

\begin{figure}[htb]
\centering
\includegraphics*[width=0.5\textwidth]{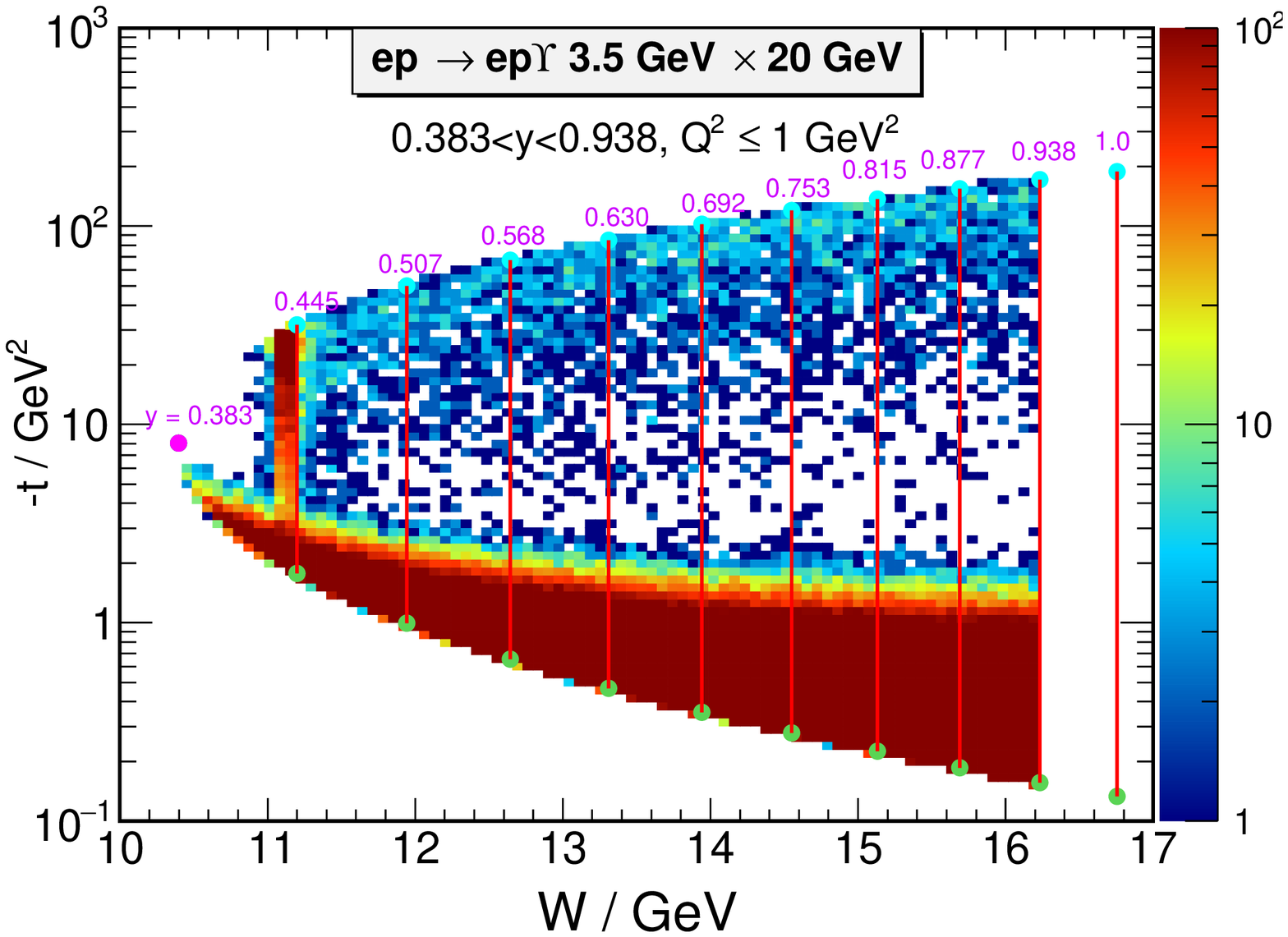}
\includegraphics*[width=0.5\textwidth]{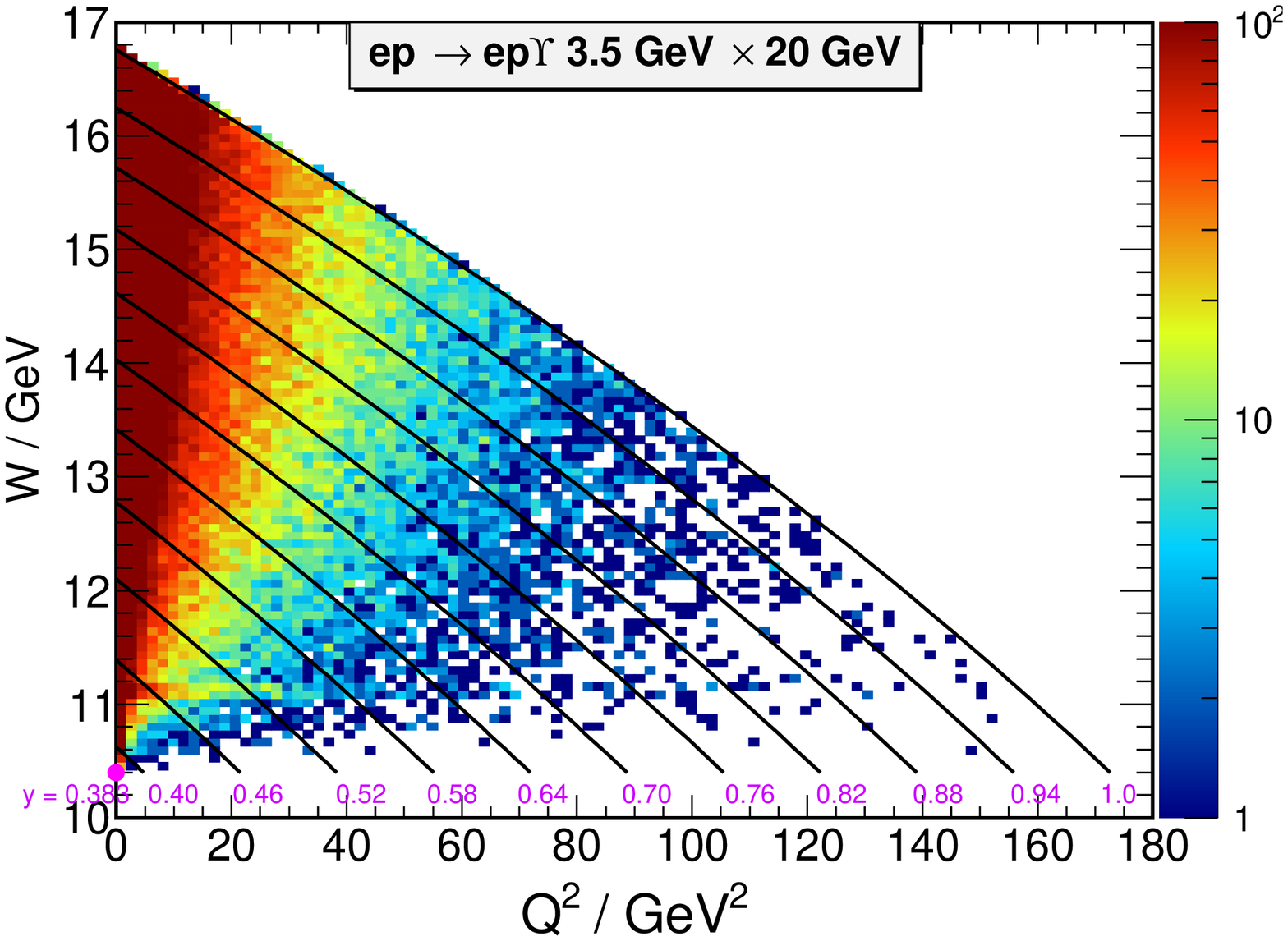}
\caption{Events of $e p \to e \Upsilon p$ in $t$-$W$ (Top) and $W$-$Q^2$ (bottom) plane under kinematical coverage of EicC. The signals of $P_b$(11120) are included in the simulation.} 
\label{fig:Upsilon}       
\end{figure}

The study of exotic particles in {the} charm sector is entering the high-precision era,
however, the pursuit of those in {the} bottom regime is just about to begin with the motivation to {derive a} complete {and unified} picture of all flavors.
According to the heavy quark flavor symmetry, the bottom
partners of exotic states in the charm sector are predicted to {safely exist}.
Up to now no {elementary} particle factory is on purpose built for exotic bottom states.
{Unlike their charm analogs, {bottom states} can hardly be produced through the weak decay of hadrons consisting of heavier quark,
{because of the very rare events of doubly- and triply-bottom hadrons and lack of the} stable top hadrons.}
Therefore, they can only be straightforwardly produced in high-energy electron-hadron and hadron-hadron collisions.
Whether there is a promising potential to observe them at an electron-ion collider
relies on the strength of {background-to-signal ratio}, both of which are of big uncertainties.
Usually the yields of the exotic bottom states are one to two orders lower than those of their {counterparts in the} Charm sectors.
A pragmatical route is to narrow the range of electroproduction cross {sections} down to 10 fb level
by the continuous improvement of accelerator luminosity and detection efficiency.
Knowing better their charm partners will certainly give impetus to this multi-flavor hunting contest.

{Attempting searches} for non-strange hidden bottom hadron resonances $P_b$ and $Z_b$ {seem to be} the most feasible {if considering} the anticipated yields.
The nominal energy of EicC is perfect for near threshold $\Upsilon$ production,
see Fig. \ref{fig:addinUpsilon} for a tentative exploration of $P_b$ via photoproduction \cite{Cao:2019gqo}.
The data of $\gamma p\to \Upsilon p$ above 100 GeV are used to estimate a safe upper limit of the photoproduction rates of non-resonant $t$-channel component, around 30 pb below $W = $ 20 GeV.
The corresponding electroproduction cross section is 0.2 pb within the coverage of low-to-medium energy electron ion colliders,
in comparison with 0.1 $\sim$ 0.2 pb if using the Pomeron or two-gluon exhange \cite{Cao:2019gqo,Wang:2019zaw}.
{Thus,} less than 10000 non-resonant events below $W =$ 20 GeV will accumulated under the integrated luminosity of 50 fb$^{-1}$,
several percentage out of which at most are possibly from the $P_b$'s decay.
As a representative 
the total spin $J$ of $P_b$ is chosen as 3/2, corresponding to the lowest orbital {momentum} $L = 0$.
The following discussion {is somewhat analogous to} the $P_b$ with spin other than $J=3/2$.
The mass $M_{P_b}  = 11.12$ GeV and width $\Gamma_{P_b} = 61.0$ MeV are used as input {to the} simulation ~\cite{Wu:2010rv,Karliner:2015voa}.
Assuming $\mathcal{B} (P_b \to \Upsilon p) = 10\%$ the peak production is around 0.3 nb as shown in Fig. \ref{fig:addinUpsilon}.
For this most optimistic case hundreds of $P_b$ would be uncovered by EIC with 50 fb$^{-1}$ run period.
However, the grey band in Fig. \ref{fig:addinUpsilon} does not consider the uncertainty of $P_b$ properties,
of which $\mathcal{B}(P_b \to \Upsilon(1S)p)$ {is especially big}.
The two-body phase space {alone} would introduce an extra reduction factor of about five for near-threshold events.
Whether it is attainable is further dependent on the machine efficiency, around 20\% $\sim$ 30\% for EicC design \cite{CAO:2020EicC,Anderle:2021wcy}.
{The reconstruction branching ratio of final $\Upsilon$ from its dilepton decay is around 2.5\% for both $e^+ e^-$ and $\mu^+ \mu^-$.}
Because of the larger mass of $\Upsilon$, the events of $\gamma^* p \to \Upsilon p$ is accumulated within {bigger range of} $Q^2$ than $J/\psi$, see Eq.~(\ref{eq:VMDformfactor}) and {the} black band in Fig. \ref{fig:addinUpsilon}.
Collecting {this} limited information at hand the exclusive $\Upsilon$ electroproduction at EicC is simulated by lAger generator in Fig. \ref{fig:Upsilon}.
The detector coverage is partly considered, but the detector resolution and reconstruction efficiency are not yet included.
The medium energy mode of US-EIC covers higher $W$ and $Q^2$ {ranges} \cite{Burkert:2022hjz}.

The yield and distribution of $Z_{b}$ events {are} similar to non-resonant $\gamma^* p \to \Upsilon p$ \cite{Cao:2020cfx}.
An additional meson {in} $\gamma p\to {B}^{*} \bar{B} \Lambda^+_b$ or $\pi^0 \Upsilon p$ needs to detected with a reduction of overall efficiency in comparison to {the} $P_b$.
The $Z_{bs}$ photoproduction rates {are} further reduced by one order at least,
{resembling} the order hierarchy of {magnitudes} between {the} $P_{c}$ ({or} $P_{b}$) and $P_{cs}$ ({or} $P_{bs}$).
The production rates of other pentaquark candidates, like ${\Lambda}_b^0$(5912) and ${\Lambda}_b^0$(5920) in the ${\gamma}p\rightarrow {\Lambda}_b^0(*)B^+$ reactions, is one to two orders lower than that of $P_b$ \cite{Huang:2021ave}.
As a consequence higher luminosity is required to investigate the electroproduction of exotic bottom states.

Several techniques would be eventually beneficial to enlarge the statistical signal of narrow and extremely heavy particles.
The missing-mass spectrum would be implemented at the expense of momentum resolution.
The {use of} light ion beam as a substitute for {the} proton beam would magnify the coherent production of states.
But incoherence from alteration of nuclear {states} may so large that the peak signal is lowered in cross sections \cite{Molina:2012mv,Paryev:2020jkp,Paryev:2020qge,Paryev:2018fyv}.
Also inclusive processes would be enhanced by several times compared to the exclusive ones in lepton-proton collisions \cite{Yang:2021jof,Winney:2022tky}.
However, pQCD calculations {are} not so optimistic, {predicting} much lower rates \cite{ColpaniSerri:2021bla}.
The actual discovery potential of exotica
states in {the} bottom sector shall be exploited under a detailed simulation of these techniques.

As previously mentioned it is practicable to search for those particles by their open bottom decay,
whose upper limit is estimated by the H1 data of $\gamma p \to b\bar{b}X$:
\bea \nonumber
   \sigma(\gamma p \to b\bar{b}X) &=&
   \begin{cases}
& 310 \pm 150 \pm 60 \pm 40 \quad \mbox{nb}  \\
& 206 \pm 19 ^{+46}_{-40} \quad \mbox{nb}
   \end{cases}
\eea
at the average of $ \langle W\rangle =$ 168 GeV ($E_\gamma = 15$ TeV) and $ \langle W\rangle =$ 180 GeV ($E_\gamma = 17.5 $ TeV), respectively~\cite{Luders:2001nb,Adloff:1999nr}.
The corresponding electroproduction cross section is:
\bea \nonumber
   \sigma(e p \to e b\bar{b}X) &=&
   \begin{cases}
& 19.5 \pm 9.3 \pm 3.7 \pm 1.8 \quad \mbox{nb}  \\
& 14.8 \pm 1.3 ^{+3.3}_{-2.8} \quad \mbox{nb}
   \end{cases}
\eea
Those data are consistent with pQCD {calculations}~\cite{Ellis:1988sb,Frixione:1994dv}.
An upper limit of photoproduction around 200 nb is given by EMC at about 20 GeV, corresponding to 1.2 pb for electroproduction \cite{EuropeanMuon:1981hfd}.
So the ratio of $\gamma p \to c\bar{c}X$ to $\gamma p \to \Upsilon p$ is nearly two orders as shown by the linear extrapolation (dash-dotted line) in Fig. \ref{fig:addinUpsilon}, which is of the same order gap between $\gamma p \to c\bar{c}X$ and $\gamma p \to J/\psi p$~\cite{CAO:2020EicC,Gryniuk:2016mpk}.
Provided that all excited open bottom states finally decay to $\bar B^{(*)}\Lambda_b$, it is foreseeable that the open bottom channels are expected to have larger cross section than that of hidden bottom ones.

Admittedly the extrapolation from the data at high energies in Fig. \ref{fig:addinUpsilon} is a very rough estimation of the non-resonant contribution at low-to-medium energies.
More quantitatively, the open bottom decays of $P_b$ are calculated by models~\cite{Xiao:2013jla,Huang:2015uda,Huang:2018wed,Lin:2018kcc} and those of $Z_b$ are rarely studied.
The ratio of $\mathcal{B}(P_b \to \bar B^{(*)}\Lambda_b)$ to $\mathcal{B}(P_b \to \Upsilon p)$ {centers} around 1.0 in quark delocalization color screening model~\cite{Huang:2018wed}, and ranging from 200 $\sim$ 1500 in the hadronic molecules picture~\cite{Lin:2018kcc}. Because the {present experiments imply} that $\mathcal{B}(P_b \to \Upsilon p) $ is smaller than 5.0\%, the latter one seems to be favored in order to saturate the total width of $P_b$
in the hypothesis of negligible decay to merely light quarks states.
So the photoproduction cross section is:
\bea \nonumber \label{eq:sigmaopen}
  \sigma_R  &\propto& \mathcal{B}(P_b \to \gamma p)\, \mathcal{B}(P_b \to \bar B^{(*)} \Lambda_b)
\eea
{where the magnitude} depends critically on the level of $\mathcal{B}(P_b \to \gamma p)$, thereby
$\mathcal{B}(P_b \to \Upsilon p)$ if VMD is retained.
Assuming {that} the decay channels with only light quarks states to be negligible, if $\mathcal{B}(P_b \to \Upsilon p)$ and $\mathcal{B}(P_b \to \eta_b p)$ are both 5.0\%, while $\mathcal{B}(P_b \to \bar B \Lambda_b)$ and $\mathcal{B}(P_b \to \bar B^{*}\Lambda_b)$ are both 47.5\%, the peak cross section of the $P_b$ is around 0.1~nb in each hidden bottom channel, and 4.75~nb in each open bottom channel. If $\mathcal{B}(P_b \to \Upsilon p)$ and $\mathcal{B}(P_b \to \eta_b p)$ is both 1.0\%, while $\mathcal{B}(P_b \to \bar B \Lambda_b)$ and $\mathcal{B}(P_b \to \bar B^{*}\Lambda_b)$ are both 49.0\%, the peak cross section of the $P_b$ is around 0.004~nb in each hidden bottom channel, and 0.196~nb in each open bottom channel. So it is really optional to detect the signals by the open bottom decay channels at the cost of reconstruct efficiency, though polarization measurement is yet unfeasible.

More optimistically, measurements of {the} bottom production near threshold,
{even though some of them may be of low statistics.}
would probe a rich of physics besides exotic bottom spectrum.
Several topics are extremely intriguing and also critically {correlated}.
It is recognized for a long time that the exclusive near-threshold photoproduction of heavy quarkonium allows for the study of the quarkonium-nucleon interaction dominated by hard gluon exchange.
Those gluon{ic processes} are due to the heavy charm or bottom quarks, thus providing a unique probe to study the gluon component in the nucleon at high $x$.
The TOTEM collaboration at the LHC and the D{\O} collaboration at the Tevatron collider at Fermilab have discovered an elusive $C$-odd state of three gluons, also known as the odderon \cite{TOTEM:2020zzr}.
The photoproduction of $C$-even quarkonium $\eta_{c(b)}$ {is} proposed as an ideal process to probe the existence of such $t$-channel exchange of a colorless gluonic compound.
The $\eta_{b}$ (or $\eta_{c}$) cross section is at the same level with $\Upsilon$ (or $J/\psi$), while $\chi_{c(b)J}$ are around one order lower in the context of {non-relativistic} QCD factorization \cite{Jia:2022oyl}.

On the other hand the photoproduction of $C$-odd heavy quarkonium in high energy {reactions} is proposed
as a way to measure gluonic densities dominated by the two gluon exchange \cite{Brodsky:1994kf,Collins:1996fb,Koempel:2011rc}.
Approaching to the threshold region, 
exclusive production of those heavy quarkonium in electron-proton scattering {may} shed light on the origin of the proton mass via the QCD trace anomaly \cite{Kharzeev:1995ij,Kharzeev:1998bz},
though it is controversial in {which} way and to what precision.
If the QCD factorization and multipole expansion remain valid in the threshold region,
 the threshold data can be interpreted in terms of the gluonic {gravitational form factor} (GFF)of the proton \cite{Guo:2021ibg}.
At large photon virtualities and very large $t$ region it is possible to
extract the gluon $D$-term in the GFF of the proton and also
probe the trace anomaly effect of QCD at the subleading level \cite{Boussarie:2020vmu}.
If holographic gauge/string duality is {considered}, the differential cross section at small $t$ region seems to be a sensitive probe of the structure of the QCD trace anomaly characterized by $b$ parameter \cite{Hatta:2018ina,Hatta:2019lxo}.
However, a perturbative QCD analysis at large-$t$ found no direct connection between the near threshold heavy quarkonium photoproduction and the gluonic GFF \cite{Sun:2021pyw,Sun:2021gmi}.
A precise {measurement} of $t$-dependent cross section near threshold would be definitely illuminating
because of the different power behavior of $t$ predicted by theories \cite{Lee:2020iuo}.

\begin{figure}[ht]
\centering
\includegraphics*[width=0.45\textwidth]{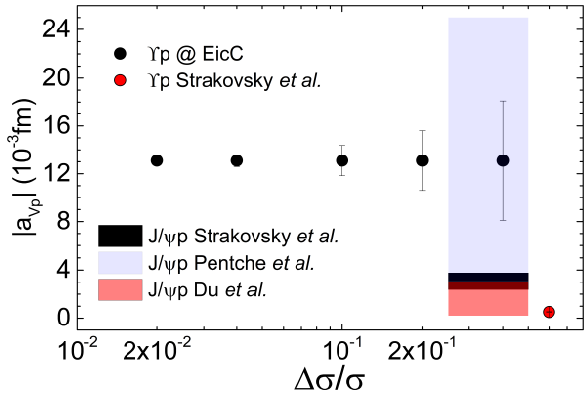}
\caption{ Expected accuracy of $\Upsilon p$ scattering length as a function of the errors of the total cross section of $\gamma p \to \Upsilon p$, confronting the centeral value (black dots) with that from Strakovsky \textit{et al.} (red dot) \cite{Strakovsky:2021vyk}.
The $J/\psi p$ scattering length extracted from GlueX data \cite{Ali:2019lzf} is shown for comparison in different schemes considering inelastic channel by Du \textit{et al.} \cite{Du:2020bqj}, and implementing VMD by
Strakovsky \textit{et al.}~\cite{Strakovsky:2019bev} and Pentchev \textit{et al.} \cite{Pentchev:2020kao}.  }
\label{fig:lengtherr}       
\end{figure}

Another enlightening topic relying heavily on VMD is relating the near threshold behavior to the quarkonium-nucleon elastic scattering length.
It is found that the kinematic corrections to standard vector dominance formulas are important long before \cite{Barger:1975ng}.
Though the extracted scattering length 0.1 $\sim$ 1 fm appears to be reasonable for light vector meson, the feasibility of this method for heavy quarkonium is never examined.
It results into a provoking order hierarchy of the absolute value of scattering length $a_{Vp}$ \cite{Strakovsky:2019bev,Strakovsky:2020uqs,Pentchev:2020kao}:
\be
|a_{\Upsilon p}| \ll |a_{J/\psi p}| \ll |a_{\phi p}| \ll |a_{\omega p}|
\ee
with $|a_{\Upsilon p}| = (0.51 \pm 0.03) \times 10^{-3}$ {fm} \cite{Strakovsky:2021vyk}
if total cross section of $\gamma p \to \Upsilon p$ is estimated {by} QCD factorization \cite{Guo:2021ibg},
and that of $\gamma p \to J/\psi p$ {is evaluated in} a VMD formalism incorporating dispersion relation  \cite{Gryniuk:2016mpk,Gryniuk:2020mlh} or three gluon-exchange model \cite{Brodsky:2000zc}.
However, we found instead a centeral value $|a_{\Upsilon p}| = 13.1 \times 10^{-3}$ {fm} in a soft dipole Pomeron model \cite{Martynov:2001tn,Martynov:2002ez}, whose error depends on the measured accuracy of cross section simulated in Fig. \ref{fig:Upsilon}.
This features the {magnitudes of scattering lengths} in another sequence:
\be
|a_{\Upsilon p}| \sim |a_{J/\psi p}| \ll |a_{\phi p}| \ll |a_{\omega p}|
\ee
fairly holding up given the large uncertainties as shown in Fig. \ref{fig:lengtherr}.
The rescattering from inelastic $\bar{D}^{(*)} \Lambda_c$ channel implies even smaller $|a_{J/\psi p}|$ \cite{Du:2020bqj}.
The $\eta_c N$ and $J/\psi N$ are both weakly attractive at short distances from quenched lattice QCD \cite{Kawanai:2010ev}, indicating the potential binding of charmonium with the nucleon and nuclei \cite{Gryniuk:2016mpk} and {moderate} interactions with hadronic matter \cite{Lenske:2018bgr}.
A small positive or negative scattering length {indicates} a weakly attractive or repulsive $J/\psi N$ interaction if there is no $J/\psi N$ bound state.
A $10^{-3}$ {fm} level of $|a_{J/\psi p}|$  and $|a_{\Upsilon p}|$ hints at that the interaction range is smaller than a typical size of a hadron and the proton is nearly transparent for heavy quarkonium.
For the purpose of quantifying the violation of VMD, it would be useful to confront those values with solid theoretical calculation.
Based on this observation a measurement of near threshold production of heavy quarkonium is definitely of its own importance.


\section{Summary and Perspective}
\label{sec:conclusion}

{At an early} era of {the} quark model,
mesons are proposed as Bosonic particles constituted by one pair of quark-antiquark ($q \bar{q}$),
and those of exotic nature are composed of two or more pairs of $q \bar{q}$.
Baryons are nominated as Fermi particles consisting of three quarks ($qqq$),
and those mentioned as exotic baryons are made up of $qqq$ plus pairs of $q \bar{q}$.
They are all thought to be genuine object in nature.
From the $\chi_{c1}$(3872) and $D_{s0}$(2317) observed by {Belle} in 2003 to the $Z_c$(3900) discovered by BESIII and {Belle} in 2013 {and up} to $Z_{cs}$(3900) established by BESIII and $T_{cc}^{+}$(3875) by LHCb in 2021,
the candidates of narrow exotic mesons are emerging as a {widely spread} family during the past two decades.
The nominees of narrow exotic baryons discovered by LHCb, though fewer, appear as another charming family {consisting} of those from $P_c$ to $P_{cs}$ pentaquarks.
However, the beautiful and ordinary light quark {families} of exotic character {seem to be} sparsely populated in light of plenty of efforts over the past years.

Photo- and electro-production of those narrow states will firmly {confirm} them as real states with lower background than hadron collisions,
though suffering from an insufficient accuracy of the production rates estimation.
The vector-meson dominance (VMD) hypothesis is usually employed, {remaining an important tool in studies of production reactions, } but never validated in terms of heavy vector quarkonium.
{VMD indeed} provides a rather accurate prediction \cite{Biloshytskyi:2022dmo} for the two photon decay width of $X$(6900) (or $T_{\psi\psi}$), a broad structure discovered by LHCb \cite{LHCb:2020bwg} and recently confirmed by ATLAS and CMS in the $J/\psi$ -pair spectrum.
It is impossible to further scrutinize this conjecture due to the failure of searching for the bottomonium correspondence $T_{\Upsilon\Upsilon}$ decaying to $\Upsilon \mu^+ \mu^-$ \cite{CMS:2020qwa}.
Nevertheless it seems worthwhile getting serious about {doubts on the validity of its predicted momentum dependence \cite{Xu:2021mju}.
In this respect, VMD should be scrutinized critically on the quantitative level by detailed studies of scattering lengths or/and radiative decay widths.}
A highlight measurement of exclusive $J/\psi$ and $\Upsilon(1S)$ near threshold production
will constitute the first clear clarification of those topics besides searching for multi-quark partners.
These measurements are seemingly feasible at the planned EicC with the design beam mode of 3.5(e) $\times$ 20(p) GeV and luminosity of $2 \sim 4 \times 10^{33}$ cm$^{-2}$ s$^{-1}$.
The {beam} polarization can be reached as high as 80\% for electron beam and 70\% for proton beam.
The light ions beam is also accessible, e.g. $^3$He beam with effective 40 GeV energy.
A low energy mode of US-EIC enables the collision of 5(e) $\times$ 41(p) GeV  with the same polarization {but} a lower luminosity \cite{Burkert:2022hjz}.
It is encouraging to pin down the quantum numbers of exotic or new hadrons of high production rates by utilization of polarized beams.
Observation of heavy exotic states of low statistics is viable through the inclusive final states at those colliders,
ultimately leading to the discovery of true hidden bottom $P_b$ states by distinguishing this benchmark resonance from kinematic{al} enhancements.
Especially, the observed production rates in different channels are closely related to the nature of the $P_b$, e.g. molecular or tetraquarks or other compact states. 

In return, a full understanding of charm and bottom exotic family will help to resolve the mystery that light quark hadrons of narrow width are all accommodated within constituent quark model patterns after taking into account the coupled-channel effect.
The understanding of the incredible absence of narrow exotic partners in light quark sector
would to large extent improve our knowledge of the nucleon {representing} the dominant parts of {the} visible universe.


\section*{\acknowledgementname}

The accomplishment of this document has benefitted from input from
many members of the EicC community, with special thanks to Kuang-Da Chao, Ling-Yun Dai, Feng-Kun Guo, Yu-Tie Liang, Qin-Yong Lin, Xiang Liu, Peng Sun, Jia-Jun Wu, Ju-Jun Xie, Ya-Ping Xie, De-Liang Yao, Zhi Yang, Zhiwen Zhao and Bing-Song Zou.
It is grateful to Horst Lenske for collaboration in Sec. \ref{sec:light} and a proofreading throughout the manuscript.
This work was supported by the National Natural Science Foundation of China (Grants Nos. 12075289 and U2032109) and the Strategic Priority Research Program of Chinese Academy of Sciences (Grant NO. XDB34030301) .

\begin{appendices}
%
\begin{figure}
  \begin{center}
  {\includegraphics*[width=0.45\textwidth]{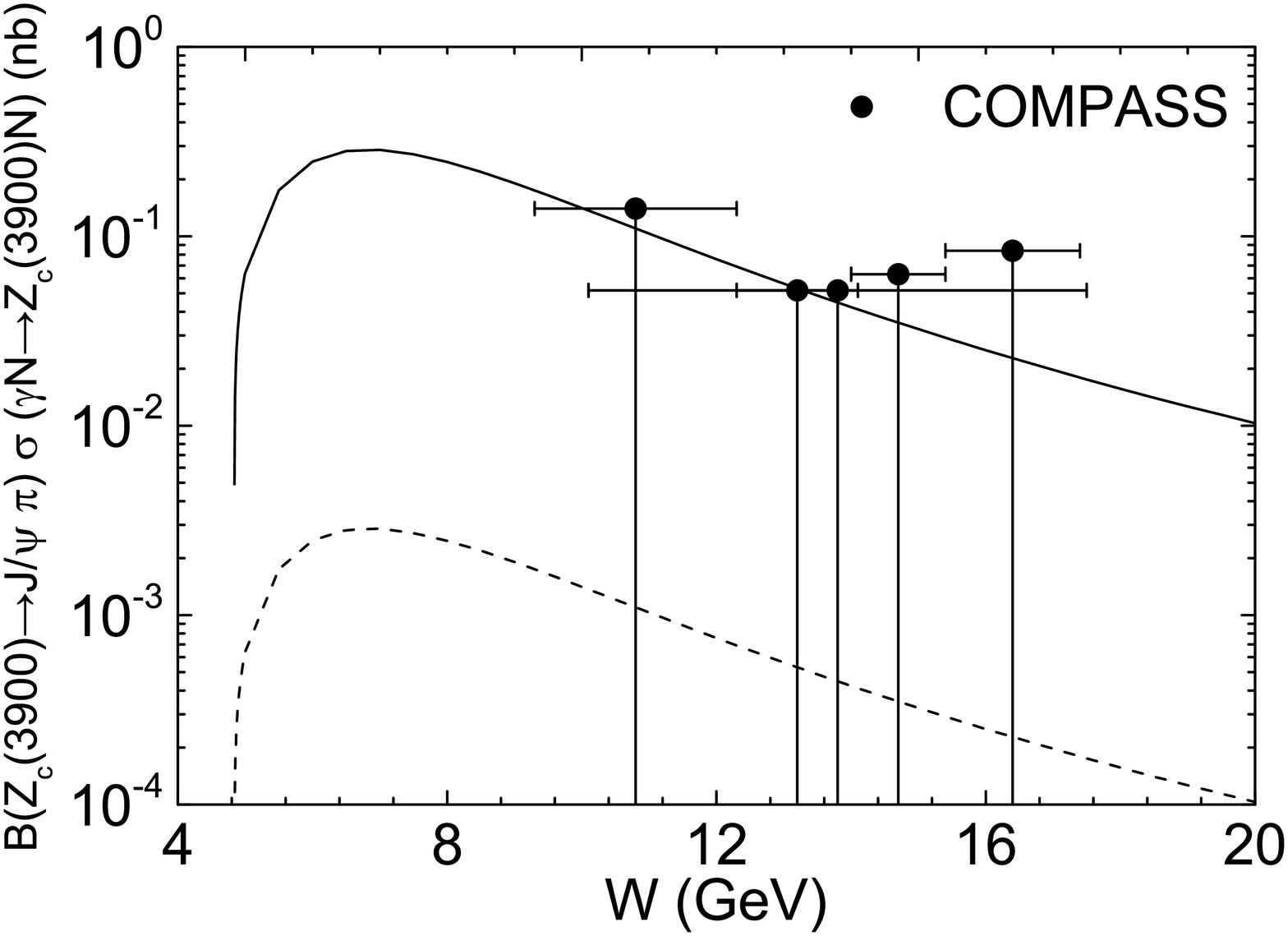}}
    \caption{The upper limit (solid curve) and lower limit (dashed curve) of cross section of $\gamma p \to Z_c^+(3900) n$. The parameterization includes $t$-channel exchange with form factor in two-body phase space \cite{Lin:2013mka}. The data is the upper limit from COMPASS Collaboration~\cite{Adolph:2014hba}.}
    \label{fig:photoZc3900}
  \end{center}
\end{figure}

\section{: About the photoproduction cross sections} \label{apdx:model}

The measurement of COMPASS collaboration gave the upper limits with 90\% confidential level at $W$ = 13.7 GeV~\cite{Aghasyan:2017utv,Adolph:2014hba}:
\bea \nonumber
\sigma (\gamma p\rightarrow \chi_{c1}(3872) p) \times  \mathcal{B}(\chi_{c1}(3872) \to J/\psi \pi^+\pi^-) &&\\ \nonumber < 2.9 \quad \mbox{pb} &&\\ \nonumber
\sigma(\gamma N\rightarrow Z_c^+(3900) N) \times \mathcal{B}(Z_c^+(3900) \to J/\psi \pi^+) &&\\ \nonumber < 52.0 \quad \mbox{pb}&&
\eea
those of $Z_c^+(3900)$ is in the same order as those of ordinary charmonium. As a result, the upper limits of electroproduction of $Z_c^+(3900)$ is in the same order as those of ordinary charmonium.
Note that we simulate the processes:
\bea \nonumber
\sigma (e p \to e \chi_{c1}(3872) p) \times \mathcal{B}(\chi_{c1}(3872) \to J/\psi \pi^+\pi^-) &&\\ \nonumber
\sigma (e p \to e Z_c^+(3900) n) \times \mathcal{B}(Z_c^+(3900) \to J/\psi \pi^+) &&
\eea
where $\chi_{c1}(3872)$ and $Z_c^+(3900)$ are reconstructed by $J/\psi \pi^+\pi^-$ and $J/\psi \pi^+$ decay, respectively. So the values of $\mathcal{B}(\chi_{c1}(3872) \to J/\psi \pi^+\pi^-)$ and $\mathcal{B}(Z_c^+(3900) \to J/\psi \pi^+)$ is not {required to be known as} a prerequisite. If assuming a reasonable upper limit of $\mathcal{B}(Z_c^+(3900) \to J/\psi \pi^+)< 100\%$, the calculated cross sections is compatible with above COMPASS data~\cite{Lin:2013mka}.

Furthermore, for the $\gamma p\to J/\psi p$ and $\gamma p\to \Upsilon p$ process in Fig. \ref{fig:channels}, the light vector meson in VMD model, e.g. $V'=\rho/\omega/\phi$, is possibly important~\cite{Wu:2019adv}. This uncertainties in VMD model is also present in above exotic meson photoproduction in $t$-channel. {The} $f(Q^2)$ depends on a mass scale $M_{\rho,\omega}$ of light vector meson:
\be \nonumber
f(Q^2) = \left( \frac{M_{\rho,\omega}^2}{M_{\rho,\omega}^2 + Q^2} \right)^2
\ee
or other similar form factors~\cite{Wu:2019adv}. Obviously above $f(Q^2)$ is dependent strongly on $Q^2$. However, to date a detailed exploration of the validity of VMD model in heavy quarkonium is not available, so this model uncertainty is not considered here.
It is also noted in the main text that that the electroproduction cross sections is insensitive to the choice of ${f(Q^2)}$, because the photon flux $\Gamma \propto Q^{-2}$ determines that most of the events spread out below $Q^2 < 1.0 $ GeV$^2$.
The model precision is adequate for simulation at this stage.

\section{: About the branching ratios of $B \to K + X/Z$ decay} \label{apdx:Bdecay}

\begin{table*}
  \caption{The branching ratios of $B \to K + X/Z$ from Particle Data Group (PDG)~\cite{ParticleDataGroup:2022pth}.
  $^\dagger$: $\mathcal{B}(\chi_{c1}(3872) \to J/\psi \pi^+ \pi^-)= (4.1^{+1.9}_{-1.1})\%$, $\mathcal{B}(\chi_{c1}(3872) \to J/\psi \gamma) = (1.1^{+0.6}_{-0.3})\%$, and $\mathcal{B}(\chi_{c1}(3872) \to D^0 \bar{D}^{*0}) = (52.4^{+25.3}_{-14.3})\%$ are used~\cite{Li:2019kpj}.
  $^\ddag$: Unknown for $Z_c(3900)$.
  $^\S$: Unknown for $Z_c(4430)$.
    \label{Tab:BtoKX}}
  \begin{center}
  \begin{tabular}{|c|c|c|c|}
 \hline
 \multicolumn{2}{|c|}{$B^+ (\times 10^{-3})$}         & \multicolumn{2}{c|}{$B^0 (\times 10^{-3})$}  \\
 \hline
  $K^+ J/\psi$    &  1.010$\pm$0.028  &  $K^0 J/\psi$     &  0.873$\pm$0.032 \\
 \hline
  $K^+ \eta_c$    &  1.09$\pm$0.09    &  $K^0 \eta_c$     &  0.79$\pm$0.12   \\
 \hline
  $K^+ \chi_{c0}$ &  0.149$\pm$0.015  &  $K^0 \chi_{c0}$  &  0.111$\pm$0.024 \\
 \hline
  $K^+ \chi_{c1}(3872)$   &  0.19$\pm$0.06$^\dagger$    &  $K^0 \chi_{c1}(3872)$    &  $0.11^{+0.05}_{-0.04}$$^\dagger$ \\
 \hline
  $K^+ Z_c(3900)$ &    ?$^\ddag$       &  $K^\pm Z_c(3900)^\mp$  & ?$^\ddag$ \\
  \qquad $\to \eta_c \pi^+ \pi^-$     & $<$0.047  &  \qquad $\to J/\psi \pi^\mp$  & $<$0.0009\\
 \hline
  $K^+ Z_c(4430)$ &    ?$^\S$          &  $K^\pm Z_c(4430)^\mp$  & ?$^\S$ \\
  \qquad $\to J/\psi \pi^+$   & $<$0.015  &  \qquad $\to J/\psi \pi^\mp$   & $0.0054^{+0.0040}_{-0.0012}$\\
  \qquad $\to \psi(2S) \pi^+$ & $<$0.047  &  \qquad $\to \psi(2S) \pi^\mp$ & $0.060^{+0.030}_{-0.024}$ \\
 \hline
 \end{tabular}
  \end{center}
\end{table*}

The branching ratios of $B \to K + X/Z$ from Particle Data Group (PDG) \cite{ParticleDataGroup:2022pth} are summarized in Tab.~\ref{Tab:BtoKX}. Note that it usually lists the product of
\be \nonumber
\mathcal{B}(B \to K + X/Z) \times \mathcal{B} (X/Z \to \mbox{final states})
\ee
To know $\mathcal{B}(B \to K + X/Z)$, it needs $\mathcal{B} (X/Z \to \mbox{final states})$. The absolute decay branching ratios of $\chi_{c1}(3872)$ are just recently measured by Babar group~\cite{BaBar:2019hzd} and consistent with a full analysis~\cite{Li:2019kpj}. The deduced $\mathcal{B}(B \to K + \chi_{c1}(3872))$ is smaller than $\mathcal{B}(B \to K + J/\psi)$ and $\mathcal{B}(B \to K + \eta_c)$ by a factor of 5$\sim$10, and in the same level of $\mathcal{B}(B \to K + \chi_{c0})$, see Tab.~\ref{Tab:BtoKX}. The absolute decay branching ratios of $Z_c^+(3900)$ and $Z_c^+(4430)$ is not fixed so far. If their branching ratios of $\eta_c \pi^+ \pi^-$, $J/\psi \pi^+$ and $\psi(2S) \pi^+$ are smaller than 10\%, then $\mathcal{B}(B \to K + Z_c)$ is only smaller than $\mathcal{B}(B \to K + J/\psi)$, $\mathcal{B}(B \to K + \eta_c)$ and $\mathcal{B}(B \to K + \chi_{c0})$ by a factor of around ten.
In the simulation, it {is} reasonable to assume very roughly a factor of 100 reduction of cross section of exotic mesons in comparison of conventional charmonium states, based on the experience from $B$-meson decay, see dashed curve in Fig. \ref{fig:photoZc3900}.

\end{appendices}

%
%
%

\end{document}